\documentclass{jfm-for-arXiv}
\usepackage{graphicx}
\usepackage{epstopdf, epsfig}

\usepackage{bm}
\usepackage{graphicx}
\usepackage{amsmath}
\usepackage{amsfonts}
\usepackage[usenames,dvipsnames]{xcolor}
\usepackage{float}
\usepackage{tabularx}
\usepackage{wrapfig}
\usepackage{natbib}

\usepackage{subcaption}
\usepackage{hyperref}
\usepackage{tcolorbox}

\usepackage{comment}

\newcommand*\Kai[1]{\textcolor{Red}{#1}}

\newcommand*\Bruce[1]{\textcolor{Plum}{#1}}

\shorttitle{Singularity Formation in 2D Euler Equations}
\shortauthor{J. Bergmann, T. Maurel--Oujia, X.--Y. Yin, J.--C. Nave and K. Schneider}

\title{Singularity formation of vortex sheets in 2D Euler equations using the characteristic mapping method}

\author{Julius Bergmann\aff{1,2}
  \corresp{\email{julius.bergmann .at. univ-amu.fr}},
  Thibault Maurel--Oujia\aff{1},
  Xi--Yuan (Bruce) Yin\aff{3},
  Jean--Christophe Nave\aff{4}
 \and Kai Schneider\aff{1}}

\affiliation{
\aff{1}Institut de Mathématiques de Marseille, Aix--Marseille Université, CNRS, Marseille, France
\aff{2}Institut für Strömungsmechanik und Technische Akustik, Technische Universität Berlin, Berlin, Germany
\aff{3}LMFA--CNRS, Ecole Centrale de Lyon, Lyon, France
\aff{4}Department of Mathematics and Statistics, McGill University, Montréal, Québec, Canada}

\begin{document}

\maketitle

\begin{abstract}
%Abstract   
The goal of this numerical study is to get insight into singular solutions of the two-dimensional (2D) Euler equations for non-smooth initial data, in particular for vortex sheets. To this end high resolution computations of vortex layers in 2D incompressible Euler flows are performed using the characteristic mapping method (CMM). This semi-Lagrangian method evolves the flow map using the gradient-augmented level set method (GALS). The semi-group structure of the flow map allows its decomposition into sub-maps (each over a finite time interval), and thus the precision can be controlled by choosing appropriate remapping times. Composing the flow map yields exponential resolution in linear time, a unique feature of CMM, and thus fine scale flow structures can be resolved in great detail. Here the roll-up process of vortex layers is studied varying the thickness of the layer showing its impact on the growth of palinstrophy and possible blow up of absolute vorticity. The curvature of the vortex sheet shows a singular-like behavior. The self-similar structure of the vortex core is investigated in the vanishing thickness limit. Conclusions on the non-uniqueness of weak solutions of 2D Euler for non-smooth initial data are drawn and the presence of flow singularities is revealed tracking them in the complex plane.\\
\end{abstract}

\begin{keywords}

\end{keywords}

%-----------------------------------------------------------------------------------
\section{Introduction}
\label{sec:intr}

\bigskip

The emergence of coherent vorticity is ubiquitous in high Reynolds number turbulence, as for example in shear layers or in wall bounded flow in form of thin boundary layers detaching and traveling into the bulk flow. %maybe some refs.
The dynamics of thin vortex layers in the limit of vanishing viscosity is of tremendous interest for understanding the formation of singularities in the incompressible 2D Euler equations with non-smooth initial data.
In contrast to 2D Euler with smooth initial conditions, for which results on global regularity and uniqueness of the solution are well known \citep{Yudovich1963}, it was shown by \citet{Szekelyhidi2011} that for vortex sheet initial data infinitely many non-stationary weak solutions exist, which moreover conserve energy.
For a review on the mathematics of turbulence we refer to \citep{Majda2002,Bardos2013} and for 2D vortex layers we refer to the detailed introduction of \citet{Caflisch2022}.

In a recent work \citet{Caflisch2022} showed by means of direct numerical simulation using a classical pseudo-spectral method that 2D vortex layers may have complex singularities and after the roll up of the layer that the vortex cores may have unbounded vorticity in the limit of infinite Reynolds numbers.
We recall that \citet{Caflisch2022} did almost exclusively viscous computations by solving Navier--Stokes using Fourier pseudo-spectral methods and coupling the vortex layer thickness with the viscosity, i.e. the inverse Reynolds number. Some computations for the inviscid case, i.e. for 2D Euler, were likewise presented in \citet{Caflisch2016} and \citet{Caflisch2022}.
However, the inviscid computations were limited to minimum layer thickness values ($\delta=0.0141$). Note that smaller values %of $\delta$ 
did not work due to the necessary resolution requirements, however they also used larger values (cf. table 3 in \citet{Caflisch2022}).
The motion of an inviscid vortex sheet, i.e. for vanishing thickness, 
is governed by the Birkhoff--Rott model equation \citep{Birkhoff1962, Rott1956} which develops a singularity in finite time starting from smooth initial data \citep{Moore1979}. Perturbations grow due to the Kelvin--Helmholtz instability and the vortex sheet does roll up. Regularized simulations using a vortex blob method have been also performed in \citet{Caflisch2022} and compared with Navier--Stokes computations.

{The vanishing viscosity limit of 2D Navier--Stokes in the presence of boundaries was likewise studied in \citet{nguyen2011energy, nguyen2018energy}. Dipole-wall collisions were simulated and the existence of a Reynolds-independent energy dissipation rate was shown.
In this context Prandtl’s  classical  boundary  layer argument was complemented, which states that both the boundary-layer thickness and dissipation rate are proportional to $Re^{-1/2}.$
However for detaching boundary layers, Kato’s scaling was shown to be more appropriate than Prandtl’s scaling, which implies that the boundary layer scales with $Re^{-1}$.
Some reviews of 2D flows with walls can be found in \citet{clercx2017dissipation} and mathematical analysis of weak solutions of the 2D Navier--Stokes equations in bounded domains, in the vanishing viscosity limit, in \citet{constantin2019vorticity}.
}

State of the art for solving Navier--Stokes or Euler equations numerically with high precision are pseudo-spectral methods \citep{HuZa1987, IsGK2009} which  have been also extensively used for investigating nearly singular solutions of the 3D Euler equations \citep{HoLi2007, GiBu2008}.
A Cauchy--Lagrange method for computing 2D Euler flows was proposed in \citet{Podvigina2016}. This semi-Lagrangian method exploits the time-analyticity of fluid particle trajectories and was shown to be more efficient than pseudo-spectral computations.
However, detailed singularity studies have not been reported so far.
An even more powerful tool for solving the incompressible Euler equations is the Characterisitic Mapping Method (CMM).
Evolving the flow map with a Gradient Augmented Level Set Method, developed in \citep{Nave2009, seibold2011jet, chidyagwai2011comparative}, one can decompose the long time deformation into short time sub-maps due to the semi-group structure of the flow map. This yields a numerical scheme with exponential resolution in linear time developed for linear advection in \citet{mercier2020characteristic} and 2D Euler in \citep{CMM2D}, allowing to capture the exponential growth of vorticity gradients.
The implementation of the method has global third order convergence in space and time and its efficiency has been demonstrated in comparison with spectral and Cauchy--Lagrange methods in \citep{CMM2D}. More recently an extension to 3D incompressible Euler flows has been proposed in \citep{CMM3D}.
The compositional adaptivity of CMM is an essential feature which allows detailed insight into the small scales of the solution without using prohibitive numerical resolutions.

The goal of the present paper is to revisit the vortex layer computations of \citet{Caflisch2022} in the inviscid case in more depth using CMM and to extend the range to smaller $\delta$-values and longer times. As presented in \citet{Caflisch2016}, for too thick vortex layers the vortex merging process starts to interfere with the formation of the two vortex blobs with spiral arms.
Here we compute solutions of 2D incompressible Euler flows and study the dynamics of these extremely thin vortex layers in the vanishing thickness limit and investigate possible singularities. The aim is to get some insights into the non-uniqueness of week solutions and about possible singularities. Curvature and vortex strength of the vortex center-line are analyzed and a temporal and spatial normalization unveils the dynamics for vanishing thickness limit, as well as the investigation of singularities in the complex plane. The palinstrophy growth and energy spectra show and distinguish the impact of both the forming vortices and vortex merger process. Thanks to the high resolution capabilities of CMM we get insight into the fine scale structure of vortex cores and their dynamics in Euler flows.

The remainder of the paper is organized as follows. Set-up and initial conditions are discussed in section~\ref{sec:set-up}. A short description of the characteristic mapping method for solving the 2D incompressible Euler equations is given in section~\ref{sec:cmm}. Section~\ref{sec:perfomredcomputations} gives an overview on the performed computations and numerical results are then presented in section~\ref{sec:results}. A singularity analysis in the complex plane is performed in section~\ref{sec:singularityanalysis}.
Finally, conclusions and perspectives for future work are given in section~\ref{sec:concl}.

%-----------------------------------------------------------------------------------
\section{Governing equations and initial condition}
\label{sec:set-up}

\begin{figure}
    \begin{subfigure}[t]{0.48\linewidth}
        \centering
        \includegraphics[height=5.3cm]{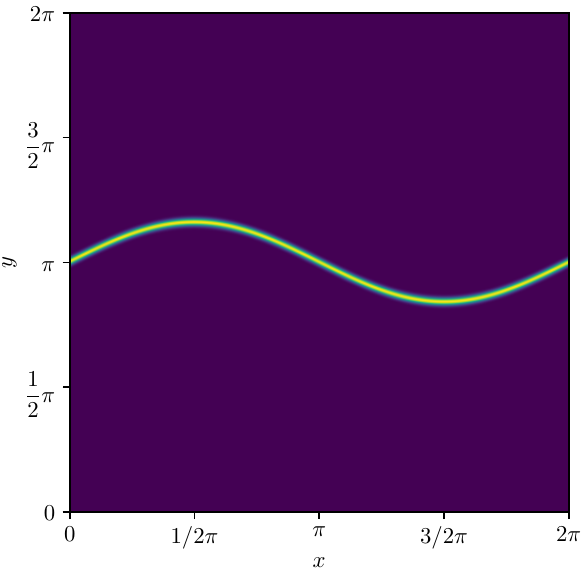}
        \includegraphics[height=5.3cm]{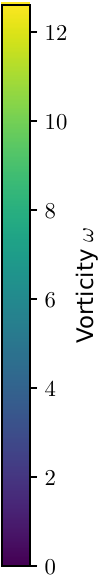}
        \caption{Initial condition $\omega_0$ for $\delta\approx 0.032$}
    \end{subfigure}
    \hspace{0.2cm}
    \begin{subfigure}[t]{0.42\linewidth}
        \centering
        \includegraphics[height=5.3cm]{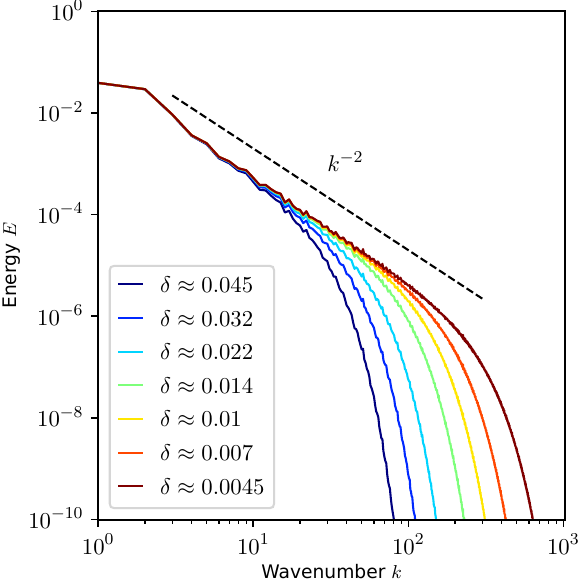}
        \caption{Energy spectrum for different $\delta$-values}
        \label{fig:IC-power-spectra}
    \end{subfigure}
    \caption{Initial condition for $\delta\approx0.032$ and energy spectra for different $\delta$-values.} 
    \label{fig:IC-vorticity}
\end{figure}

We consider inviscid flow in a $2 \pi$ periodic domain $\Omega = [ 0 , 2\pi] \times [ 0 , 2\pi]$ in the plane, governed by the 2D incompressible Euler equations. Starting point is the vorticity transport equation,
\begin{equation}
    \partial_t \omega + ( {\bm{u}} \cdot  \nabla ) \omega = 0   
    \label{eqn:euler-vorticity}
\end{equation}
where  the vorticity is defined as $\omega = \nabla \times {\bm u}$ and $\bm{u}$ is the incompressible  velocity, satisfying $ \nabla \cdot \bm{u} = 0 $.
The curl operator is invertible and the velocity can be computed from the vorticity, $\bm{u} = - \nabla \times \Delta^{-1} \omega$
using the Biot--Savart operator.
Equation~\ref{eqn:euler-vorticity} is completed with a suitable initial condition $\omega_0({\bm{x}}) = \omega({\bm{x}}, t=0)$, here a regularized vortex sheet
and where $\bm{x} = (x,y)$ with $(x,y) \in \Omega$ as the cardinal directions.
All variables have been non-dimensionalized, similar to \citet{Caflisch2022}, as follows,
\begin{subequations}
\noindent\begin{tabularx}{\linewidth}{@{}XXXX@{}}
\begin{equation}
    \bm{x} = \bm{x}^*\frac{1}{\lambda}
\end{equation} &
\begin{equation}
    t = t^* \frac{\Gamma}{\lambda^2}
\end{equation} &
\begin{equation}
    \bm{u} = \bm{u^*}\frac{\lambda}{\Gamma}
\end{equation} &
\begin{equation}
    \bm{\omega} = \bm{\omega^*}\frac{\lambda^2}{\Gamma}
\end{equation}
\end{tabularx}
\end{subequations}
With the domain length $L_x = L_y = 2 \pi$ we have for the spatial scaling $\lambda = L_x/(2 \pi) = 1$. The circulation $\Gamma$ equals $2 \pi$, as shown below.
The regularized vortex sheet initial condition proposed in \citet{Caflisch2022} reads,
\begin{align}
    \omega_0(x,y) =  \frac{1}{\sqrt{2 \pi} \delta}\exp \left( -\frac{(y - \pi - \phi(x))^2}{2 \delta^2}\right)
    \label{eqn:vort-init}
\end{align}
where $\delta$ is the thickness parameter and $\phi$ a perturbation function. 
The initial field corresponds to a vorticity line with Gaussian cross section of thickness $\delta$ centered around a perturbation function $\phi(x)$ shown in figure \ref{fig:IC-vorticity}. The profile for $\phi(x)$ is sinusoidal with $\phi(x) = \sin(x)/2$ and $(x,y) \in \Omega$. The position of the center line $\phi$ is the curve of maximum vorticity, which oscillates around $y=\pi$. For $\delta \rightarrow 0$ this vortex layer converges towards a vortex sheet used in \citet{Caflisch2022} as initial condition for the Birkhoff--Rott equation. \\
The energy spectrum of the initial condition describes the initial frequencies present in the flow field. It is defined as:
\begin{align}
    E(k,t) = \frac{1}{2} \sum_{k-1/2 \le |{\bm k}| < k+1/2} |\widehat {\bm u}({\bm{k}, t})|^2 \label{eqn:energy-spectra}
\end{align}
Here $\widehat \cdot$ denotes the 2D Fourier transform.
The initial condition, a vortex line regularized with a Gaussian cross section, does in the limit of $\delta \rightarrow 0$ approach a vortex line, i.e. a Dirac distribution. The enstrophy spectrum thus exhibits a  $k^{0}$ scaling and consequently the energy spectrum yields a $k^{-2}$ scaling, using the relation $E(k,t) = k^{-2} Z(k,t)$.
For large $\delta$ values the regularization becomes more and more visible resulting in a faster, i.e. exponential decay for large wave-numbers, as illustrated in figure~\ref{fig:IC-power-spectra}.
For viscous flow simulations, considered in \citet{Caflisch2022}, $\delta$ was related via $\delta= Re^{-1/2}$, where $Re= \Gamma / \nu$ is the circulation based Reynolds number with $\Gamma = \int \omega({\bm{x}}, t=0) d {\bm{x}}$ being the initial circulation of the vortex layer and $\nu$ being the kinematic viscosity. Note that in the present study $\nu$ vanishes. 
For the given initial condition re-shifted to $\Omega_s = [-\pi, \pi] \times [-\pi, \pi]$, the initial circulation is
\begin{align}
    \int_{\Omega_s} \omega_0(\bm{x}) d\bm{x} & = \int_{-\pi}^{\pi}  \int_{-\pi}^{\pi} \frac{1}{\sqrt{2\pi}\delta} \exp{\left(-\frac{(y-\phi(x))^2}{2 \delta^2}\right)} dx dy  \\ 
    & = \int_{-\pi}^{\pi} \int_{(-\pi - \phi(x))/2\delta}^{(\pi - \phi(x))/2\delta} \frac{1}{\sqrt{\pi} } \exp{\left(-v^2\right)} dv dx \nonumber \\
    & = \int_{-\pi}^{\pi} \frac{1}{2} \Bigl[ {\rm{erf} } (v) \Bigr]_{(-\pi - \phi(x))/2\delta}^{(\pi - \phi(x))/2\delta} d x = 2 \pi + o (\exp ( - \delta^{-2}) ) . \nonumber
\end{align}

The evaluation limits $(\pm \pi - \phi(x) ) / 2\delta$ arise from the truncation of the Gaussian profile by the periodic box (on the upper and lower sides of which there is technically a discontinuity). 

Here ${\rm erf}(z) = \frac{2}{\sqrt{\pi}}\int_{0}^{z} \exp(-v^2) dv$ is the error function which rapidly approaches $1$ for $z \rightarrow \infty$, i.e. for $\delta \rightarrow 0$. Hence the error function can be approximated by $1$ for all investigated thickness values $\delta$ to a certainty far below machine precision, justified by the expansion ${\rm erf}(x) = 1 - e^{-x^2} \frac{1}{\sqrt \pi} \left( \frac{1}{x} - \frac{1}{2 x^3} + \frac{3}{4 x^5} - \frac{15}{8 x^7} \right) + o (x^{-8} e^{-x^2})$ \citep{Abramowitz1968}. The term $[ {\rm{erf} } (v) ]_{(-\pi - \phi(x))/\delta}^{(\pi - \phi(x))/\delta}$ can therefore approximated to 2 with error of order $o( \exp(-(\pi - \frac{1}{2})^2 \delta^{-2}) )$ using that $| \phi(x) | \leq \frac{1}{2}$.

\begin{figure}
    \centering
    \includegraphics[height=5.3cm]{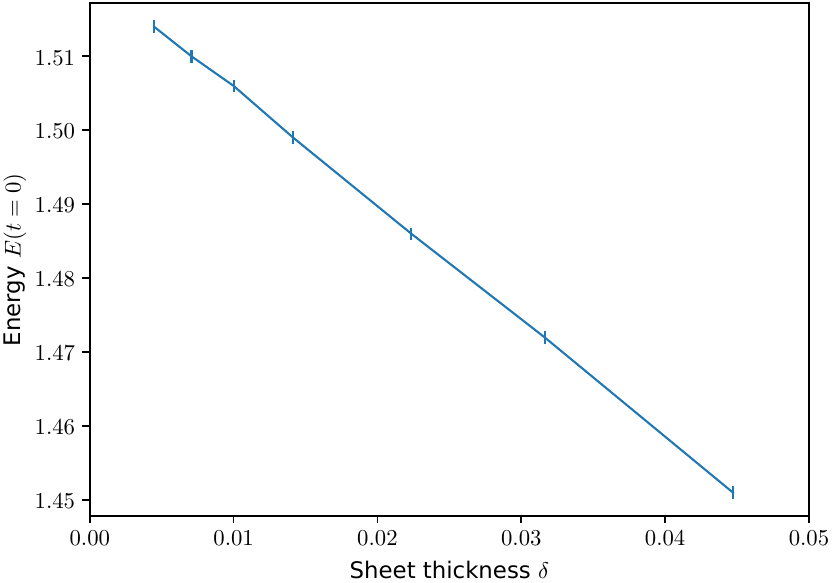}
    \caption{Initial energy $E(t=0)$ as a function of the sheet thickness $\delta$ showing a linear decrease with a value of $\approx 1.52$ for $\delta \rightarrow 0$.}
    \label{fig:IC-Energy-Comp}
\end{figure}

For incompressible and inviscid flow, energy $E(t)$ and enstrophy $Z(t)$ remain constant over time while the  palinstrophy does increase super exponentially \citep{Ayala2014}. Those quantities are defined via
\begin{align}
    E(t) &= \frac{1}{2} \int_{\Omega} |\bm{u}(\bm{x},t)|^2 d\bm{x} \, , \\
    Z(t) &= \frac{1}{2} \int_{\Omega} |\omega(\bm{x},t)|^2 d\bm{x} \, ,\\
    P(t) &= \frac{1}{2} \int_{\Omega} |\nabla \omega(\bm{x},t)|^2 d\bm{x} \, .
\end{align}
With varying thickness $\delta$, the initial energy $E(t=0)$ was found to increase linearly with decreasing vortex sheet thickness with a limit value of $\approx 1.52$ for $\delta \rightarrow 0$ (figure \ref{fig:IC-Energy-Comp}), 
The initial enstrophy $Z(t=0)$ is shown below to scale with $\delta^{-1}$ and the initial palinstrophy $P(t=0)$ with $\delta^{-3}$, both values thus go to infinity with vanishing vortex sheet thickness.
Similar to the circulation, for the shifted domain $\Omega_s$ we get for the initial enstrophy:
\begin{align}
    \frac{1}{2} \int_{\Omega_s} |\omega_0(\bm{x})|^2 d\bm{x} & = \frac{1}{2}\int_{-\pi}^{\pi}  \int_{-\pi}^{\pi} \frac{1}{2\pi\delta^2}\exp{\left(-\frac{(y-\phi(x))^2}{\delta^2}\right)} dx dy \\
    &= \frac{1}{2} \int_{-\pi}^{\pi} \int_{(-\pi - \phi(x))/\delta}^{(\pi - \phi(x))/\delta} \frac{1}{2\pi \delta } \exp{\left(-v^2\right)} dv dx  \nonumber \\
    &= \frac{1}{2} \int_{-\pi}^{\pi} \frac{1}{2 \sqrt{\pi} \delta } \Bigl[ {\rm{erf} } (v) \Bigr]_{(-\pi - \phi(x))/\delta}^{(\pi - \phi(x))/\delta} dx = \frac{\sqrt{\pi}}{2 \delta }  + o (\exp ( - \delta^{-2}) ) . \nonumber 
\end{align}

Respectively, the initial palinstrophy can be computed analytically with the gradient of the initial vorticity:
\begin{equation}
\nabla \omega_0 = - \left( \begin{matrix}
    - \phi'(x) / \delta \\ 1 / \delta
\end{matrix}  \right)  \frac{y - \phi(x)}{\delta}   \omega_0 (x, y)
\end{equation}
We thus obtain
\begin{align}
    \int_{\Omega_S} | \nabla \omega_0 |^2 d \bm{x} & = \frac{1}{2 \pi \delta^3}\int_{-\pi}^\pi \int_{(-\pi - \phi(x))/\delta}^{(\pi - \phi(x))/\delta} \left( 1 + (\phi'(x))^2 \right) v^2 \exp(-v^2) d v d x \\
    & = \frac{1}{8 \pi \delta^3} \int_{-\pi}^\pi  \left( 1 + \frac{\cos^2(x)}{4} \right)   \Bigl[  \sqrt{\pi} {\rm{erf}} (v) - 2v \exp( -v^2)  \Bigr]_{(-\pi - \phi(x))/\delta}^{(\pi - \phi(x))/\delta}  dx  \nonumber \\
    & = \frac{9 \sqrt{\pi}}{32 \delta^3} + o ( \exp(- \delta^{-2} ) ). \nonumber
\end{align}
 We bound the size of this error using $\max \phi(x)$, by $\rm{erf} (( \pi - 1/2)/\delta)  = 1 - o(\exp(-( \pi - 1/2)/\delta)^2) )$ and similarly for the $2u \exp( -u^2)$ term. This then gives the initial palinstrophy $P(t=0) = \frac{9 \sqrt{\pi}}{32 \delta^3} + o ( \exp(- \delta^{-2} ) )$.

\begin{comment}
\Bruce{For energy, I dont have a good calculation. If $\phi(x) = 0$, the velocity field is a regularized sawtooth function which goes to the below as $\delta \to 0$
\[ u^1 (x, y) = \left\{ \begin{matrix}
    \frac{y}{2 \pi} - 1/2 & y \in (0, \pi) \\
    \frac{y}{2 \pi} + 1/2 & y \in (-\pi, 0) 
\end{matrix}   \right. \]
which has kinetic energy $\pi^2/6 \approx 1.64$. If we apply the transformation $y \mapsto y + \phi(x)$ directly to $u^1$, we get the same vorticity, but the transformed velocity is no longer divergence-free and there is a pressure term to be subtracted. The resulting kinetic energy will be lower, but I couldn't directly compute the $L^2$ of that pressure gradient.} \Kai{Ok, so let's use Julius' extrapolation.}
\end{comment}

The numerical values for normalized initial enstrophy $Z(t=0) \cdot \delta = \frac{\sqrt{\pi}}{2} \approx 0.886 $ and palinstrophy $P(t=0) \cdot \delta^3 = \frac{9\sqrt{\pi}}{32} \approx 0.499$ encountered in the simulations are consistent with the analytically derived values up to machine precision ($10^{-16}$). The same applies for the initial circulation, matching the value $2\pi$.

\begin{comment}
\begin{table}[ht]
    \centering
    \begin{tabular}{ c || c | c | c | c | c | c | c}
        $\delta \approx$ & 0.0045 & 0.007 & 0.01 & 0.0141 & 0.022 & 0.032 & 0.045 \\ \hline
        $E(t=0)$ & 1.514 & 1.510 & 1.506 & 1.499 & 1.486 & 1.472 & 1.451 \\
        % $Z(t=0)$ & 198.2 & 125.3 & 88.62 & 62.67 & 39.63 & 28.03 & 19.82 \\
        % $Z(t=0) \cdot \delta$ & 0.886 & 0.866 & 0.866 & 0.866 & 0.866 & 0.866 & 0.866 \\
        % $P(t=0)$ & $5.6 \cdot 10 ^{7}$ & $1.4 \cdot 10 ^{6}$ & $5.0 \cdot 10 ^{5}$ & $ 1.7 \cdot 10 ^{5}$ & $4.4 \cdot 10 ^{4}$ & $1.5 \cdot 10 ^{4}$ & $5.6 \cdot 10 ^{3}$ \\
        % $P(t=0) \cdot \delta^3$ & 0.499 & 0.499 & 0.499 & 0.499 & 0.499 & 0.499 & 0.499 \\
        % $\Gamma(t=0) - 2\pi$ & $-8 \cdot 10^{-15}$ & $-5 \cdot 10^{-15}$ & $-2 \cdot 10^{-15}$ & $8 \cdot 10^{-16}$ & $ < 1 \cdot 10^{-16}$ & $8 \cdot 10^{-16}$ & $8 \cdot 10^{-16}$
    \end{tabular}
    \caption{Initial energy for different $\delta$-values.}
    \label{tab:energy}
\end{table}
\end{comment}

\bigskip

%-----------------------------------------------------------------------------------
\section{Characteristic mapping for 2D Euler}
\label{sec:cmm}

\begin{figure}
\centering
\begin{subfigure}[t]{0.3\linewidth}
\centering
    \includegraphics[width=\linewidth]{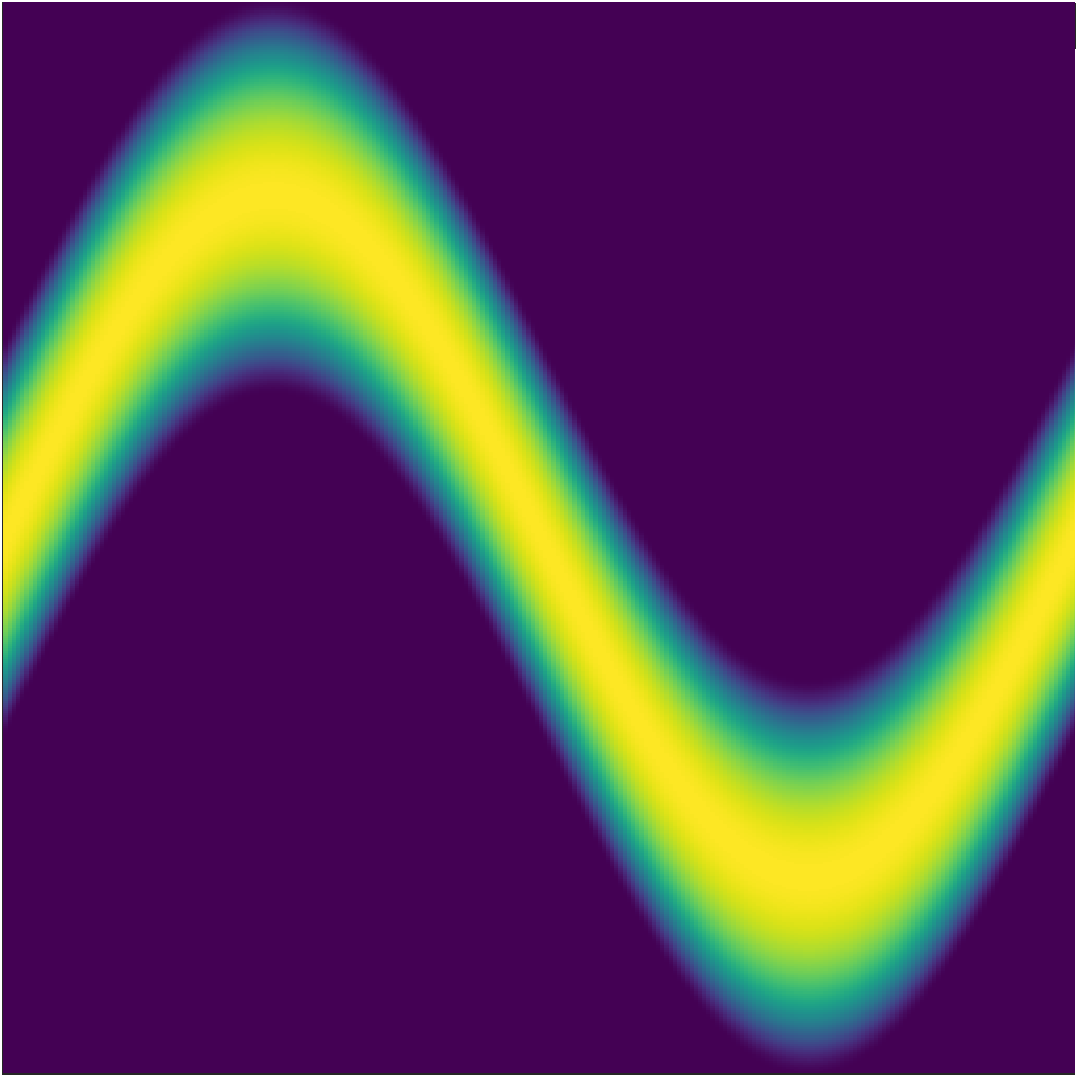}
    \caption{}
    % \caption{Initial condition of vorticity field}
\end{subfigure}
\hspace{0.2cm}
\begin{subfigure}[t]{0.3\linewidth}
\centering
    \includegraphics[width=\linewidth]{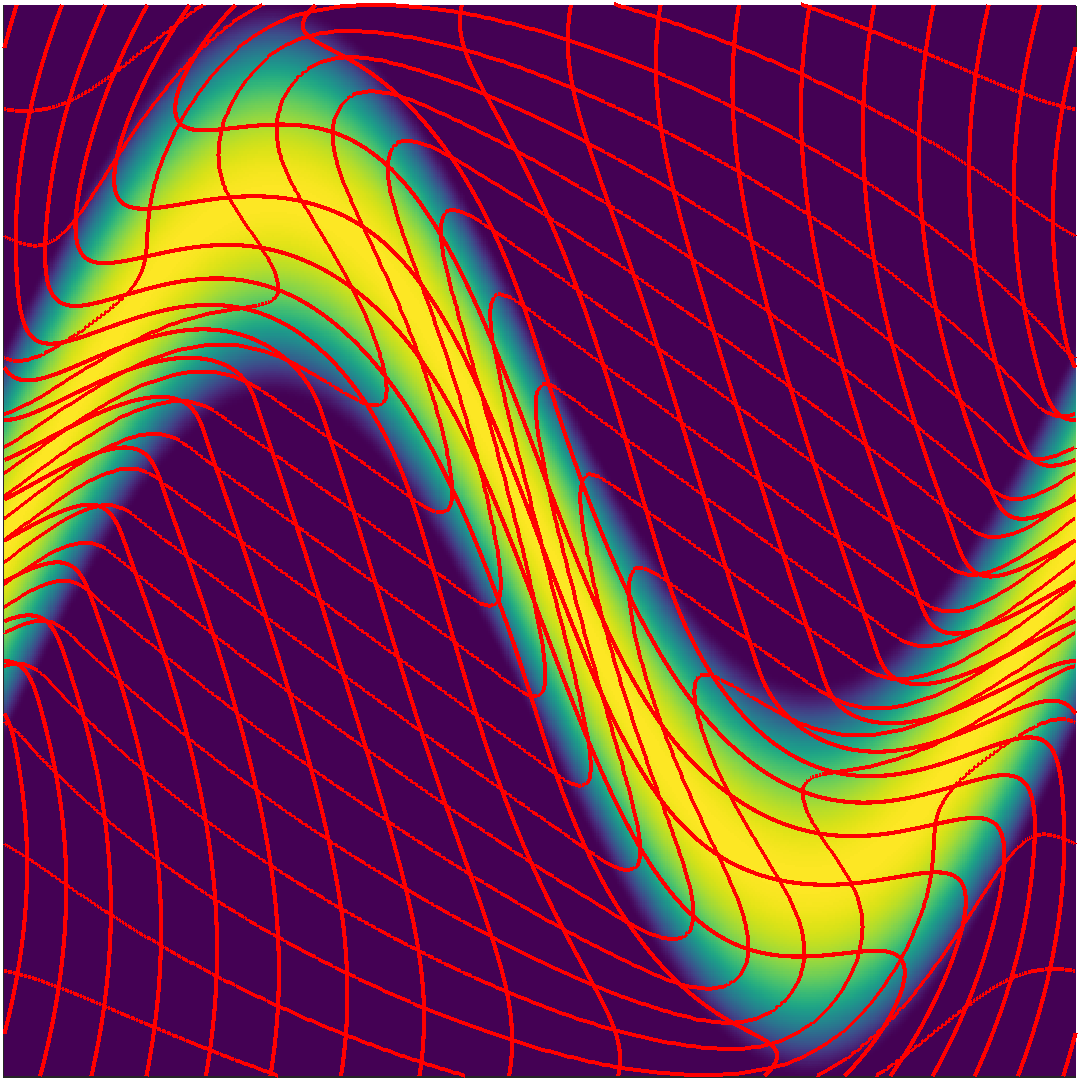}
    \caption{}
    % \caption{Relabelling symmetry using the characteristic map}
\end{subfigure}
\hspace{0.2cm}
\begin{subfigure}[t]{0.3\linewidth}
\centering
    \includegraphics[width=\linewidth]{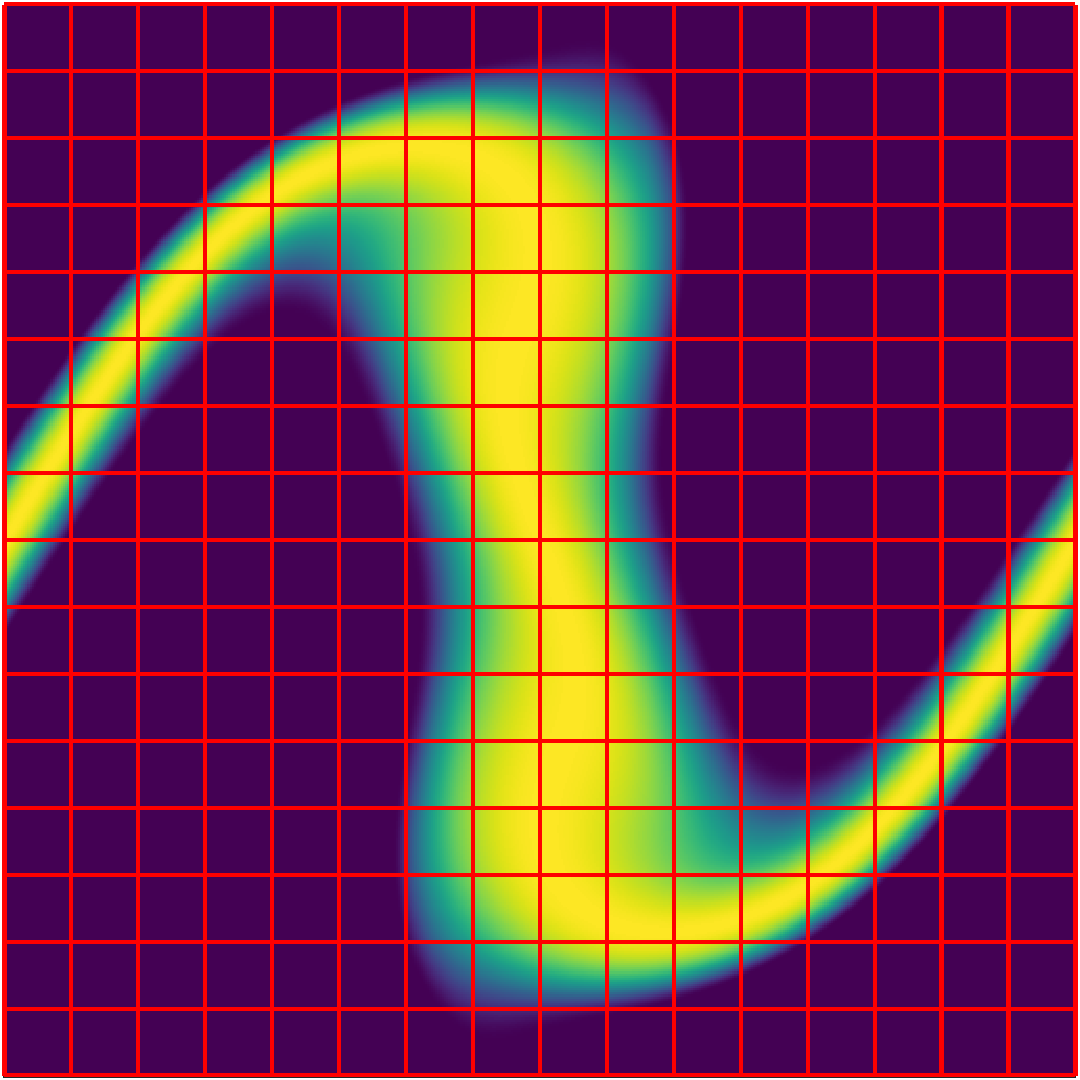}
    \caption{}
    % \caption{Solution obtained by pullback of the initial condition by the map}
\end{subfigure}
\caption{Illustration of the pullback operation by the characteristic map. Figure (a) shows the initial vorticity, figure (b) overlays onto figure (a) the backward map, and figure (c) shows the time $t$ vorticity obtained by pullback using the relabelling symmetry.}
\label{fig:IC-drawback}
\end{figure}

In this section, we briefly recall the characteristic mapping method (CMM) for solving the 2D incompressible Euler equations. For details and numerical implementation we refer the reader to \citet{CMM2D}. The numerical results in this paper are computed using a GPU Cuda implementation of the method. The open source code is available on GitHub \citep{url_cmm}.

The CMM for the 2D incompressible Euler equations is based on the 2D scalar vorticity formulation (equation \ref{eqn:euler-vorticity}) coupled with the CMM for linear transport. We recall that the characteristic map $\chi_B$ is the backward Lagrangian flow map and can be thought of as the back-to-label operator for the advection of Lagrangian particles. Indeed, for any particle trajectory $\gamma(t)$ given by,
\begin{equation}
    \frac{d}{dt} \gamma(t) = \bm{u}( \gamma(t), t)
\end{equation}
 with initial condition $\gamma(0) = \gamma_0$, we have that $\chi_B$ satisfies,
\begin{equation}
    \gamma_0 = \chi_B( \gamma(t), t) .
\end{equation}
It follows by the method of characteristics, that any scalar quantity $\phi$ advected by the same velocity $\bm{u}$ satisfies $\phi(\bm{x}, t) = \phi_0 (\chi_B(\bm{x}, t))$, this is known as the relabelling symmetry.

In the vorticity equations, $\omega$ is a scalar-valued quantity advected by $\bm{u}$ and thus has the above relabelling symmetry. The velocity field $\bm{u}$ is in turn obtained from the Biot--Savart law. % \ref{eqn:velocityDefn}. 
The coupling of the vorticity equations with the characteristic map yields the following governing equations for the numerical method:
\begin{subequations}
    \begin{align}
        & \partial_t \chi_B + ( {\bm u} \cdot \nabla ) \chi_B = 0 \label{eqn:mapEvolution} \\
        & \omega ( {\bm x} , t ) = \omega_0 ( \chi_B( {\bm x}, t)) \label{eqn:vorticityPullback} \\
        & {\bm u} = - \nabla \times \Delta^{-1} \omega \label{eqn:velocityDefn}
    \end{align}
\end{subequations}
Numerically, the advection equation for the map \eqref{eqn:mapEvolution} is discretized through the Gradient-Augmented Level-Set method \citep{Nave2009, seibold2011jet, chidyagwai2011comparative} which is a semi-Lagrangian method using Hermite cubic spatial interpolation with Runge--Kutta time-stepping schemes. The vorticity evaluation step \eqref{eqn:vorticityPullback} is performed on a fixed fine grid by interpolation of the Hermite cubic representation of $\chi_B$ followed by a direct evaluation of the initial condition $\omega_0$. Figure \ref{fig:IC-drawback} illustrates the computation of the vorticity by pullback on the initial condition. Lastly, the velocity field is defined through the Biot--Savart law \eqref{eqn:velocityDefn} as the curl of the Hermite interpolant of the stream function, which we obtain from a spectral solver for the Poisson equation using Fast Fourier Transforms. The time dependence of the velocity field is obtained from a Lagrange extrapolation in time using stream function data from the three most recent time steps.

This approach has several desirable numerical properties for the problem investigated here. Firstly, the use of the advection solution operator $\chi_B$ maintains a back-to-label symmetry of the vorticity field. This ensures a non-dissipative numerical method, since there exists a coordinate transform which transports the numerical solution to the exact one. This property is beneficial to the study of singularity formation since artificial viscosity is eliminated as potential regularization mechanism preventing blow-up. Secondly, the characteristic maps benefit from the structure of the Lie-groups of volume-preserving diffeomorphisms. Therefore, a long-time map can be represented as the composition of multiple short-time sub-maps. For a time subdivision $0 < t_1 < t_2 < \ldots < t_{m-1} < T$, the sub-map decomposition is an adaptive and multi-resolution representation of the full map $\chi_B$ given by
\begin{equation}
    \chi_B(\cdot, T) = \chi_{[t_1, 0]} \circ \chi_{[t_2, t_1]} \circ \dots \circ \chi_{[T, t_{m-1}]} \label{eqn:submapDecomp}
\end{equation}
where $\chi_{[t_i, t_{i-1}]}$ represents the backwards map in the time interval $[t_{i-1}, t_i]$. On each time sub-interval, the spatial deformation is comparatively small and can be well resolved on a coarser grid, the remapping step is triggered dynamically as the numerical resolution of the sub-map coarse grids are depleted. The full map obtained from the map composition in \eqref{eqn:submapDecomp} retains all scales formed by each sub-map. \\

The movement of the vortex sheet center line, parametrized by $\bm{x} (\theta, t)$, is tracked by following the trajectories of Lagrangian particles in the vortex core. 
The initial position corresponds to the perturbation function $\phi(x)$, i.e. $\bm{x} (\theta, t=0) = (\theta, \phi(\theta) + \pi) , \quad \theta \in [0, 2\pi] $.
The initial vortex center curve is discretized using $N_P$ uniformly distributed sample particles with $\theta_n = n \, 2 \pi / N_p$. A sufficiently high number of points $N_P \geq 10 ^6$ ensures that all dynamics of the material line is captured even under strong elongation effects. \\
Their time evolution under the numerical velocity field computed from the CMM is then tracked individually using standard explicit Runge--Kutta methods. We must note here that the velocity field obtained from the CMM is only $C^0$ in space since it is defined as the curl of the Hermite cubic interpolant of the stream function. The size of the discontinuities in higher derivatives depends on the grid size $N_{\psi}$ used for the stream function with the jumps in $i^{th}$ derivative scaling like $\| D^4 \psi \|_{\infty} N_{\psi}^{-4+i}$. This introduces a negligible error in the $L^\infty$ error for the curve position and derivatives but can be a source of noise for the regularity analysis.

%-----------------------------------------------------------------------------------

\section{Performed computations}
\label{sec:perfomredcomputations}

Several runs with successively decreasing $\delta$-values were executed on state-of-the art graphics cards, maximizing the available usable memory. All used parameters are summarized in table~\ref{tab:sim-parameters}. Two grid sizes were used: A coarser one for the description of the flow map $\chi$ and velocity $\bm{u}$ and one for the initial vorticity $\omega_0$ defined for each sub-map as well as the discrete vorticity used in the Biot--Savart law. The merit of this lies in the enhanced smoothness of the velocity field, obviating the need for a highly detailed representation. Additionally, fine scales of the flow captured by the flow map are retained from composition of the individual sub-maps. The time-step $\Delta t$ is set after a Courant--Friedrich--Lewis (CFL) number of $1/3$ of the coarse map to neglect its influence, however the semi-Lagrangian method can easily deal with large CFL numbers $>1$ as well. The incompressibility threshold of $\delta_{inc,b}$ ensures the volume preservation of the sub-maps, i.e. we monitor that $| \det \nabla  \chi_B| \le \delta_{inc,B}$. The parameter of the local stencil size $\epsilon_m$ defines the distance at which spatial derivatives for the creation of the Hermite interpolants with the GALS method are computed, together with its corresponding map update stencil order. The filter size $k_{LP}$ defines the cut-off frequency for the low-pass filter and was disabled by setting it to high frequencies of negligible energy. The fluid and embedded particle time schemes are chosen to adapted versions of the Runge--Kutta scheme with improved efficiency for the CM method. At last, $N_P$ is the number of equidistant particles embedded at the vortex center line. Further in-depth analysis and explanation of all used parameters are reported in \citet{CMM2D, Bergmann2022}. \\
\begin{figure}
	\centering % [width=0.95\linewidth]
    \begin{subfigure}[t]{0.26\linewidth}
        \centering
        \includegraphics[height=3.8cm]{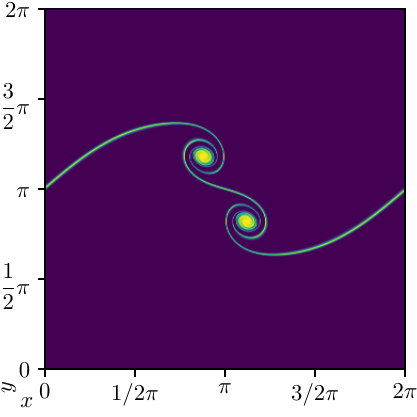}
        \caption{$t=5$ and $\delta\approx 0.032$}
    \end{subfigure}
    \hspace{0.2cm}
    \begin{subfigure}[t]{0.34\linewidth}
        \centering
        \includegraphics[height=3.8cm]{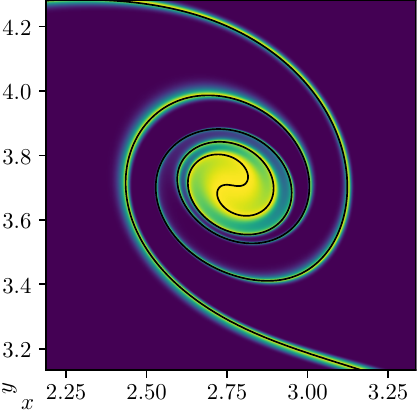}
        \includegraphics[height=3.8cm]{images/Vort-CB-delta0_032-borders.pdf}
        \caption{Close-up with overlayed center line}
        \label{fig:flow-vort-5-close}
    \end{subfigure}
    \hspace{0.2cm}
    \begin{subfigure}[t]{0.34\linewidth}
        \centering
        \includegraphics[height=3.8cm]{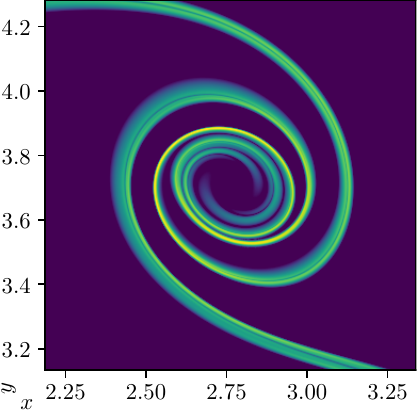}
        \includegraphics[height=3.8cm]{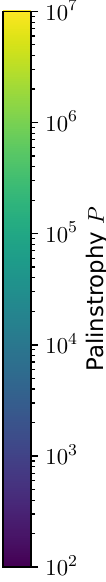}
        \caption{\centering Close-up of local palinstrophy}
        \label{fig:flow-pal-5}
    \end{subfigure}
	\caption{Developed vortices in global field and close-up of upper vortex with overlayed center-line (a,b). After an initial phase two symmetric vortices start to form with center with condensed region of high vorticity and spiral arms. Figure (c) shows corresponding local palinstrophy $|\nabla \omega(\bm{x},t)|^2$.}
	\label{fig:flow}
\end{figure}
\begin{figure}
	\centering % [width=0.95\linewidth]
    \begin{subfigure}[t]{0.45\linewidth}
        \centering
        \includegraphics[width=\linewidth]{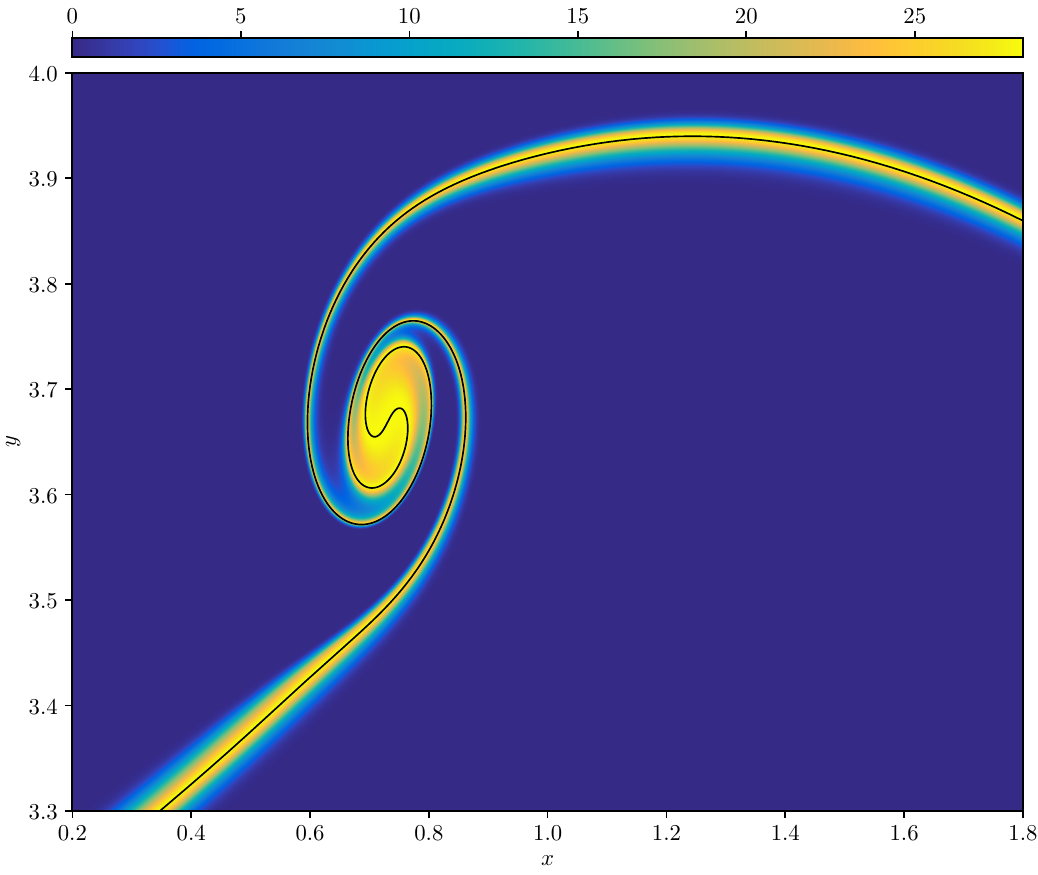}
        \caption{Results obtained with CMM} %from this paper}
        \label{fig:comp-caflisch-1}
    \end{subfigure}
    \hspace{0.2cm}
    \begin{subfigure}[t]{0.45\linewidth}
        \centering
        \includegraphics[width=\linewidth]{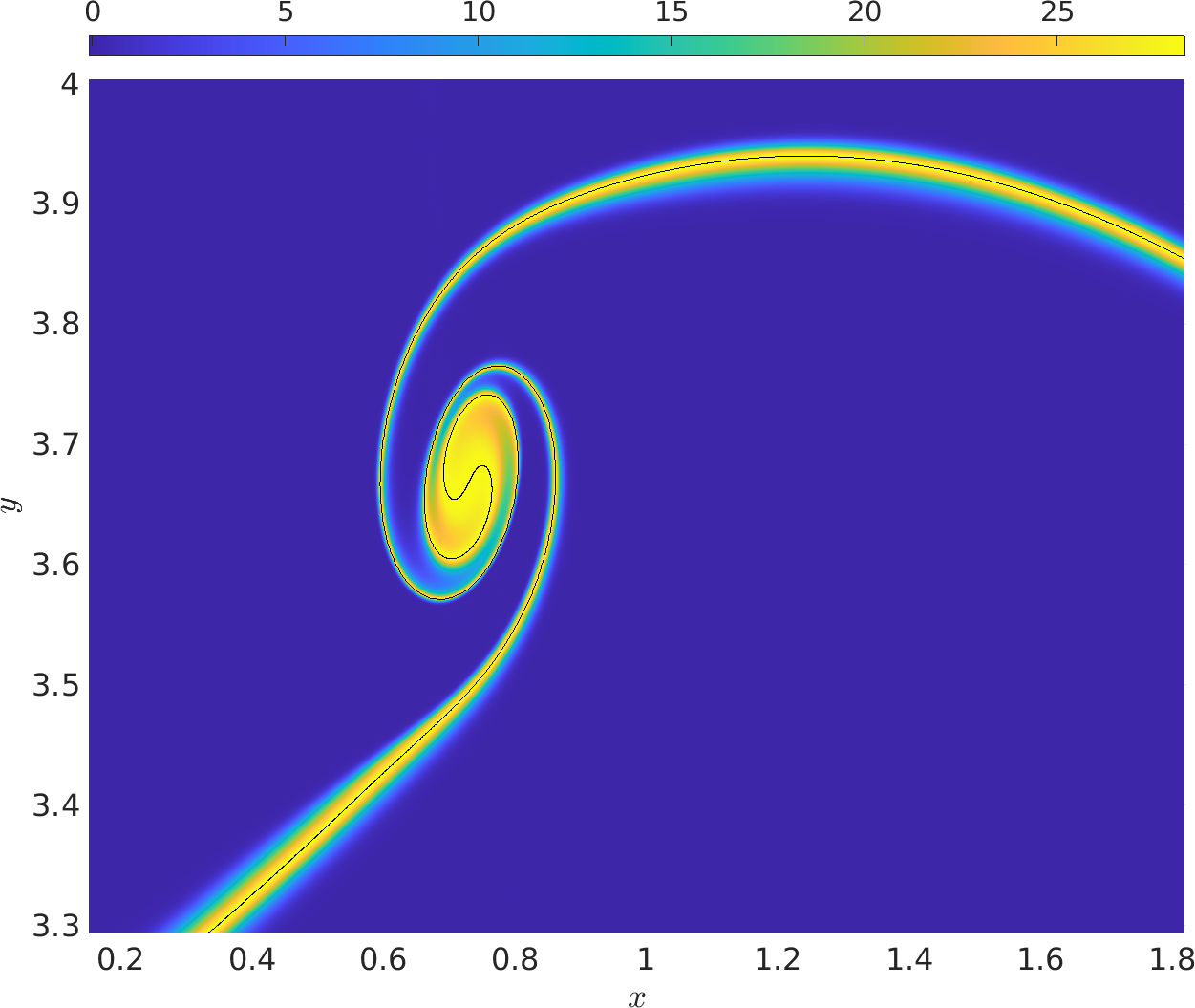}
        \caption{Results from \citet{Caflisch2022}}
        \label{fig:comp-caflisch-2}
    \end{subfigure}
	\caption{Comparison between our results and \citet{Caflisch2022} with $\delta\approx0.014$ and at $t=2.85$. The CMM results %from this paper
 were adapted to the shifted region $\Omega_s$ for comparison purposes. They show excellent agreement. Figure adapted from \citet{Caflisch2022}, printed with their permission.}
	\label{fig:compcaflisch-caflisch2022}
\end{figure}
The parameters were specifically chosen to maximize the coarse grid size for the given GPU memory, which in return leads to lower growth of incompressibility error and therefore more accurate flow representation over time. With too low grid resolution, a simulation with small thickness $\delta$ leads to premature emergence of Kelvin--Helmholtz (KH) instabilities along the line of high vorticity. This was observed to be suppressed with increased grid size and therefore attributed to numerically induced errors. For the highest available grid settings on the NVIDIA A100 cards, a value of $\delta\approx0.007$ was the lowest observed value where perturbations are bounded and do not dominate the flow behavior. In simulations with smaller vortex sheet thickness the flow is completely dominated by the emerging KH-vortices before any roll-up process starts to occur. In total, simulations for six different $\delta$-values have been performed successfully ranging from $\delta\approx0.007=1/\sqrt{2\cdot10^4}$ to $\delta\approx0.045=1/\sqrt{5\cdot10^2}$. The values were placed with $1/\delta^2$ in order to ensure comparability to the Navier--Stokes results from \citet{Caflisch2022} by their overall dynamics, as they did for their inviscid results.

\begin{table}%[ht]
    \centering
    \begin{tabular}{ c | c | c || c | c | c}
        Parameter & V100 & A100 & Parameter & V100 & A100 \\ \hline
        $N_{\text{coarse}}$ \& $N_{\psi}$  & $8192$ & $12288$ & $N_{\text{fine}}$ \& $N_{\omega}$ & $12288$ & $24576$ \\
        $\Delta t_{\text{fluid}}$ & $1/24576$ & $1/36864$ & $\delta_{inc,b}$ & $10^{-3}$ & $10^{-3}$ \\
        $\epsilon_m$ & $10^{-4}$ & $5\cdot10^{-5}$ & $k_{LP}$ & $4096$ & $12288$ \\
        Fluid time scheme & RK4Mod & RK3Mod & Map update stencil & 4th order & 4th order \\
        $N_P$ & $10^6$ & $10^7$ & Particle time scheme & RK4Mod & RK3Mod \\
    \end{tabular}
    \caption{Simulation parameters for the different runs on Nvidia V100 and A100 machines. Given are the resolution of coarse and fine grids $N_{coarse}$ and $N_{fine}$, time-step $\Delta t_{\text{fluid}}$, incompressibility threshold $\delta_{inc,b}$, local stencil size $\epsilon_m$ for the GALS method with its stencil order, low-pass filter size $k_{LP}$, time scheme order for fluid and particle advection and number of particles.}
    \label{tab:sim-parameters}
\end{table}

\begin{figure}
	\centering % [width=0.95\linewidth]
    \begin{subfigure}[t]{0.45\linewidth}
        \centering
        \includegraphics[width=\linewidth]{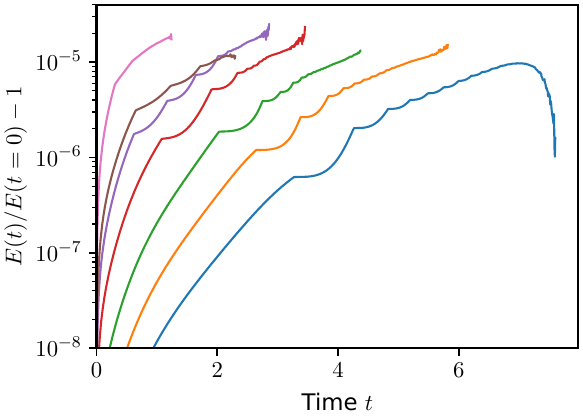}
        \caption{\centering Relative energy error $\frac{E(t)}{E(t=0)} - 1$
        }
    \end{subfigure}
    \hspace{0.2cm}
    \begin{subfigure}[t]{0.45\linewidth}
        \centering
        \includegraphics[width=\linewidth]{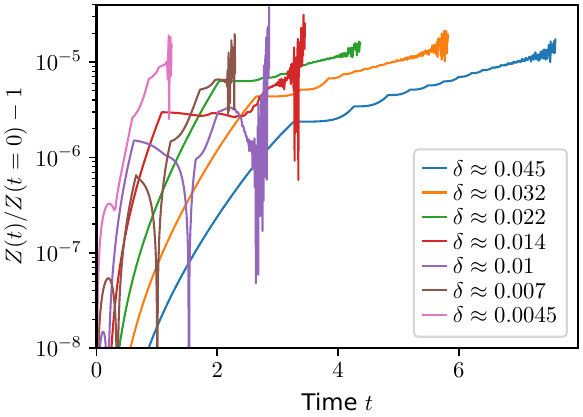}
        \caption{\centering Relative enstrophy error $\frac{Z(t)}{Z(t=0)} - 1$
        }
    \end{subfigure}
	\caption{Relative energy and enstrophy error to the initial condition $t=0$ for simulations with different $\delta$-values over time. Both quantities are well conserved up to $2\cdot10^{-5}$.}
	\label{fig:energy-enstrophy}
\end{figure}

All runs were performed until the map-stack of emerging sub-maps depleted the CPU RAM of the available compute nodes. The computations were computed on the IDRIS supercomputer. Up to 43 convoluted maps for A100 runs and 117 convoluted maps for V100 runs were captured. Details on the executed run architecture and final observed time together with the amount of captured windings of the emerging vortex structures are given in table~\ref{tab:sim-details}. The settings ensure that energy and enstrophy are conserved to an error of $\approx2\cdot10^{-5}$, see figure~\ref{fig:energy-enstrophy}, which is supported by the error of the CM-method being not dissipative in nature for transported quantities.

\begin{table} %[ht]
    \centering
    \begin{tabular}{ c || c | c | c | c | c | c | c}
        $\delta \approx$ & 0.0045 & 0.007 & 0.01 & 0.0141 & 0.022 & 0.032 & 0.045 \\
        $1/\delta^2$ & $5\times10^4$ & $2\times10^4$ & $1\times10^4$ & $5\times10^3$ & $2\times10^3$ & $1\times10^3$ & $5\times10^2$ \\
        Run on & A100 & A100 & V100 & V100 & V100 & V100 & V100 \\
        Final time & 1.24 & 2.29 & 2.86 & 3.46 & 4.37 & 5.82 & 7.59 \\
        Captured windings & 0 & 3 & 5 & 6 & 6 & 7 & 7
    \end{tabular}
    \caption{Overview on the used vorticity layer thickness values  $\delta$ including the obtained final time and the number of windings, for which computations were performed.}
    \label{tab:sim-details}
\end{table}

Simulations with successively decreasing thickness values $\delta$ reveal the flow behavior for the limit $\delta \rightarrow 0$. An example for the flow dynamics at $t=5$ and with $\delta\approx 0.032$ is depicted in figure \ref{fig:flow}. Interestingly, two small vortices start to roll up along a common center each. These have a compressed, round-shaped vortex core of high vorticity and two elongated spiral arms. Eventually, both vortices start to attract each other and will eventually merge into a single stable vortex for long times. With decreasing sheet thickness the roll-up process is triggered earlier and the emerging structures are smaller in scale. Figure \ref{fig:flow-vort-5-close} shows that the vortex core is round with distortion towards the second vortex due to the mutual attraction.
The local palinstrophy captured in figure \ref{fig:flow-pal-5}, is very pronounced in the spiral arms and shows the spiraling structures also present within the vortex core itself. \\
The acquired results are comparable with inviscid flow fields reported by \citet{Caflisch2016} and \citet{Caflisch2022}. The depicted vortex in figure \ref{fig:comp-caflisch-1} is in excellent agreement with figure 14d of \citet{Caflisch2022}, given 
in figure~\ref{fig:comp-caflisch-2}. It is important to mention that the original results are depicted in non-equal aspect ratio, due to which the vortex core appears more elliptic. \\

\begin{figure}
    \centering % [width=0.95\linewidth]
    \begin{subfigure}[t]{0.45\linewidth}
        \centering
        \includegraphics[width=\linewidth]{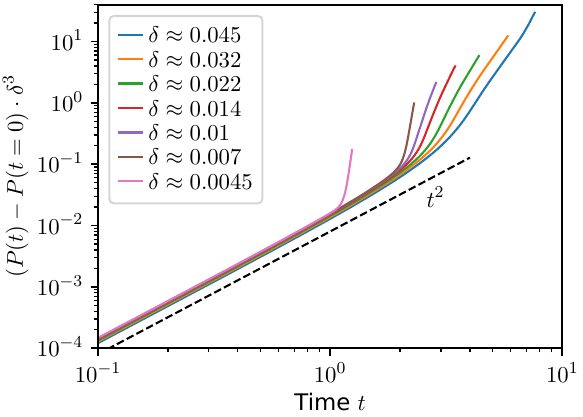}
        \caption{\centering Scaling of palinstrophy growth at early times shown in log-log.}  
        \label{fig:pal-loglog}
    \end{subfigure}
    \hspace{0.5cm}
    \begin{subfigure}[t]{0.45\linewidth}
        \centering
        \includegraphics[width=\linewidth]{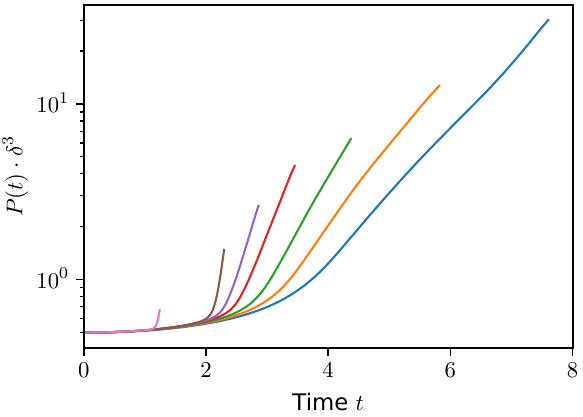}
        \caption{\centering Palinstrophy growth at later time shown in log-lin.} 
        \label{fig:pal-loglin}
    \end{subfigure}
	\caption{Initial palinstrophy growth shows a  $t^2$-scaling (a). At later times an exponential palinstrophy growth is found  which is due to the formation of rolled-up vortices (b). The steepness of the slope increases for smaller $\delta$-values and exponential growth starts at earlier times.}
	\label{fig:palinstrophy-scalings}
\end{figure}

%-----------------------------------------------------------------------------------
\section{Analysis of the results}
\label{sec:results}

In the following, the numerical results will be analyzed and discussed. Comparison of global quantities as well as the material line uncover the different flow dynamics. Normalization in space and time with respective scaling laws give insight into dynamics for the non-smooth initial condition when $\delta \rightarrow 0$. \\
The palinstrophy $P$ over time, shown in figure~\ref{fig:palinstrophy-scalings}, exhibits two different growth stages. 
At early times, all performed simulations show an algebraic growth following a $t^2$-scaling.
This initial growth period is later on overtaken by a strong exponential growth once vortex structures emerge. The steepness of the growth increases with decreasing vortex sheet thickness $\delta$.
For the slope $s$ of the exponential growth with $P(t) \propto \exp(st)$, a relation of $ s\approx \delta^{-0.77}$ was found. 
This means that in the limit of $\delta \rightarrow 0$ palinstrophy diveres.
This is consistent with results in the literature for Navier--Stokes where exponential growth of palinstrophy was derived, see e.g. \citet{lesieur2008two}. For small times a powerlaw behavior was predicted in \citet{Ayala2014} and estimates of the maximum palinstrophy growth were given.
The lower limit of the investigated $\delta$-values, i.e. for $\delta \approx 0.0045$, experiences artifacts in form of numerically accelerated secondary Kelvin--Helmholtz instabilities along the material line, resulting in the premature and irregular steep palinstrophy growth.

\begin{figure}
    \center
    \begin{subfigure}[t]{0.32\linewidth}
        \centering
        \includegraphics[height=4.1cm]{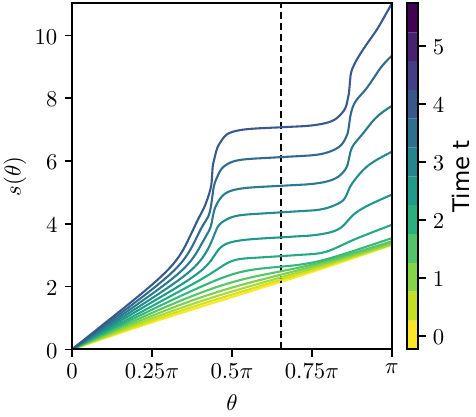}
        \caption{$s(\theta)$ for $\delta \approx 0.032$}
        \label{fig:arc-length}
    \end{subfigure}
    \hspace{0.2cm}
    \begin{subfigure}[t]{0.3\linewidth}
        \centering
        \includegraphics[height=4.1cm]{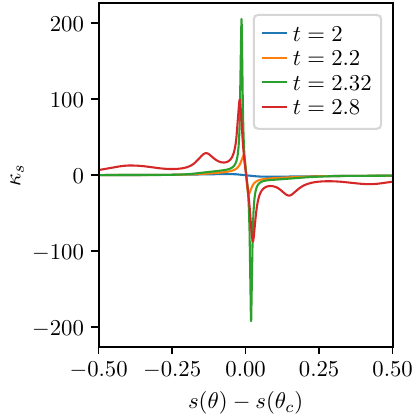}
        \caption{$\kappa_c$ for $\delta \approx 0.014$}
        \label{fig:curvature}
    \end{subfigure}
    \hspace{0.2cm}
    \begin{subfigure}[t]{0.3\linewidth}
        \centering
        \includegraphics[height=4.1cm]{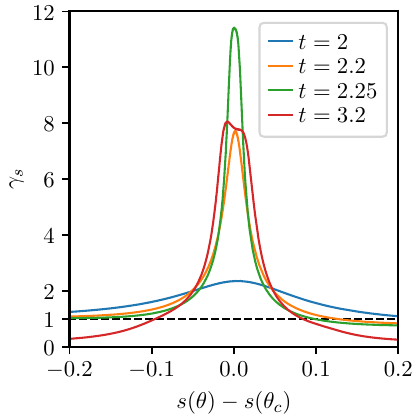}
        \caption{\centering $\gamma_c$ for $\delta \approx 0.014$}
        \label{fig:true-vortex-strength}
    \end{subfigure}
    \caption{Arc length of the material curve for $\delta = 0.032$ with vortex center as vertical dotted line (a). Curvature $\kappa_c$ (b) and true vortex strength $\gamma_c$ (b) over arc length $s(\theta)$ around vortex center for $\delta=0.014$.}
    \label{fig:material-line-quantities}
\end{figure}

With the material line $\bm{x}(\theta, t)$, 
given by the individual particle positions, we define following~\citep{Caflisch2022} several quantities, the arc length $s(\theta)$, curvature $\kappa_c(\theta)$ and true vortex strength $\gamma_c(\theta)$. These are used to analyze the material line dynamics and are defined respectively as,
\begin{align}
    s(\theta) &= \int_0^{2\pi} |\bm{x}_{\theta}| d\theta \, , \\
    \kappa_c(\theta) &= (x_\theta y_{\theta\theta} - y_\theta x_{\theta\theta}) / (x_\theta^2+y_\theta^2)^{3/2} \, , \\
    \gamma_c(\theta) &= |\bm{x}_{\theta}|^{-1} \, .
\end{align}

with the notation $\bm{x}_{\theta} = \frac{\partial \bm{x} } { \partial \theta }$ and correspondingly for the other derivatives.
Numerically the derivatives were computed using central 4th order finite differences over the particle positions for both the first and second derivative.
Both the elongation of the spiral arms and contraction around the vortex core from figure~\ref{fig:flow} can be found in the evolution of the arc length of the material line in figure~\ref{fig:arc-length}. The overall arc length is both increased by the emerging two vortices as well as the global vortex merging process with the effect intensifying strongly over time especially around the vortex core position. The almost vertical lines correspond to strong stretching of the material line, while the almost horizontal part correspond to extreme compression of the particles. Our results for the Euler case are much more pronounced but comparable with those shown in \citet{Caflisch2022} figure 12b for Navier--Stokes.

The curvature shown in figure~\ref{fig:curvature} showcases the strength of the winding process, forming two peaks of opposite sign around the vortex cores. These are situated at the edge of the coalesced vortex centers from which at longer times the spiral arms start. Involving second order derivatives the curvature experiences high-frequency oscillations from numerical derivation of the particle positions. A Gaussian filter with standard deviation of $3\cdot 10^{-4} \times N_p$ was used to mitigate this effect. In the center of the developed vortex the true vortex strength reaches a peak showcasing strong convergence towards the vortex core. Before any spiral arms are formed all particles are compressed towards the vortex core (with $\gamma_c > 1$). Once the vortices start to rotate, two spiral arms will form and elongation occurs ($\gamma_c < 1$) outside the vortex core. Both quantities reach maximum peak values at different times, depicted by the green curve in figure ~\ref{fig:true-vortex-strength}, and then start to disperse. \\
\begin{figure}
    \center
    \begin{subfigure}[t]{0.45\linewidth}
        \centering
        \includegraphics[width=\linewidth]{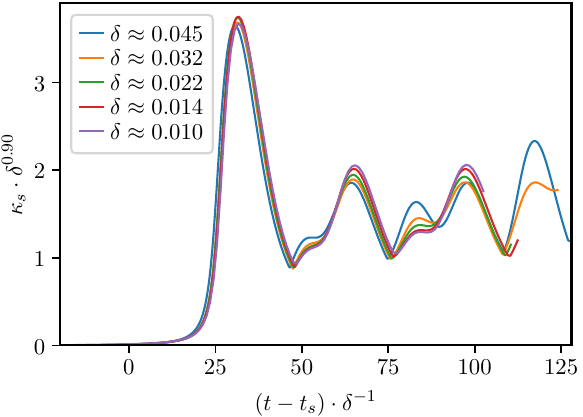}
        \caption{\centering Curvature $\kappa_c(\theta)$}
        \label{fig:curvature-rescaling}
    \end{subfigure}
    \hspace{0.2cm}
    \begin{subfigure}[t]{0.45\linewidth}
        \centering
        \includegraphics[width=\linewidth]{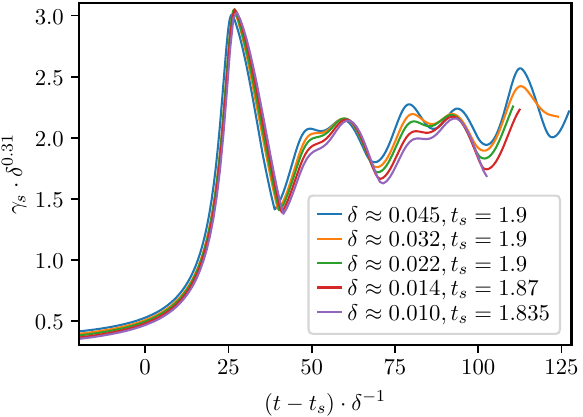}
        \caption{\centering True vortex strength $\gamma_c(\theta)$}
        \label{fig:true-vortex-strength-rescaling}
    \end{subfigure}
    \caption{Temporal re-scaling for the maximum values of the curvature $\kappa_s$ and true vortex strength $\gamma_s$ as well as temporal re-scaling for the vortex turnover with empirically determined coefficients. The critical times $t_s$, given in figure (b), were found to be decreasing with lower $\delta$-values.}
    \label{fig:rescaling-temporal-minmax}
\end{figure}
\begin{figure}
    \center
    \begin{subfigure}[t]{0.45\linewidth}
        \centering
        \includegraphics[width=\linewidth]{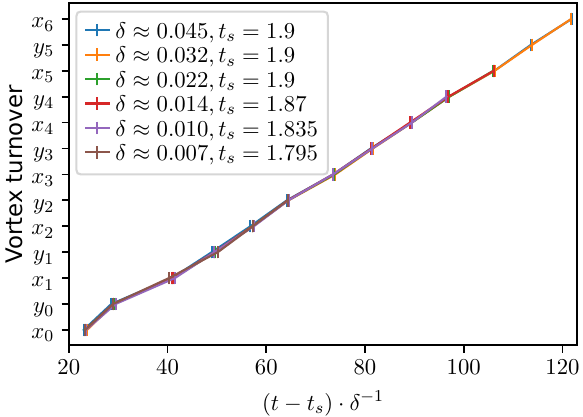}
        \caption{\centering Vortex turnover}
        \label{fig:vortex-turnover}
    \end{subfigure}
    \hspace{0.5cm}
    \begin{subfigure}[t]{0.45\linewidth}
        \centering
        \includegraphics[width=\linewidth]{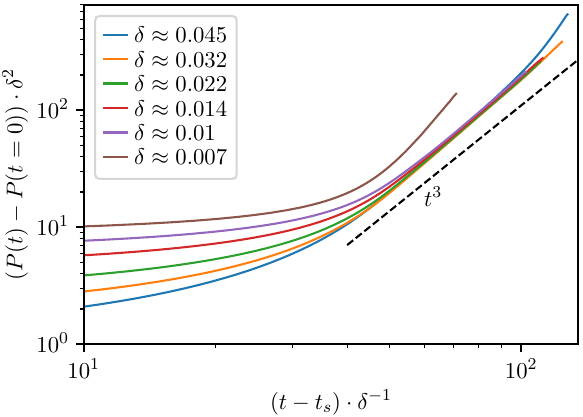}
        \caption{\centering Rescaled palinstrophy growth in log-log. }
        \label{fig:rescaling-pal}
    \end{subfigure}
    \caption{Temporal re-scaling for the vortex turnover (a) and palinstrophy growth (b) with empirically determined coefficients. The critical times $t_s$ are given in figure (a).}
    \label{fig:rescaling-temporal-2}
\end{figure}
Tracking the maximum value of curvature and vortex strength over time enables comparisons between different $\delta$-values, shown in figure~\ref{fig:rescaling-temporal-minmax}. Here, the curves have been matched empirically for the given scaling laws. The curvature was found to empirically scale with $\delta^{-0.9}$ and the true vortex strength with $\delta^{-0.31}$, for which an explanation was not found. Nonetheless, this scaling supports the conjecture of \citet{Caflisch2022} that only for the limit of $\delta \rightarrow 0$ the curvature as well as true vortex strength go towards infinity for finite time. For this limit the vortex core coalesces to an individual point with infinite curvature. \\ 
Interestingly, over time both quantities show the same behavior with an initial steep increase to an overall maximum, decreasing again with oscillations. As the observed dispersion effect appears similar for different vortex sheet thicknesses $\delta$ it is interpreted as a flow behavior rather than a result of numerical artifacts. It can be explained by the flow evolution. At first the vorticity sheet close to the vortex core starts to roll up. Two regions are formed based on the true vortex strength: One close to the center of the vortex were the material line with surrounding region of vorticity is condensed and one outside where the material line is strongly elongated from the formation of the spiral arms due to the rotation of the vortex core. At some point the vortex core is reformed into a circular shape and rotates with quasi-constant speed due to the resulting velocity from the Biot--Savart law. \\
\begin{figure}
    \center
    \begin{subfigure}[t]{0.45\linewidth}
        \centering
        \includegraphics[width=\linewidth]{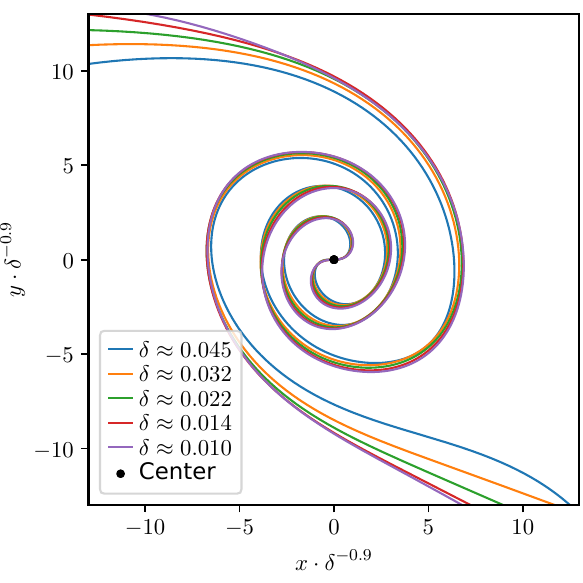}
        \caption{Re-scaled by $\delta^{-0.9}$}
    \end{subfigure}
    \hspace{0.5cm}
    \begin{subfigure}[t]{0.45\linewidth}
        \centering
        \includegraphics[width=\linewidth]{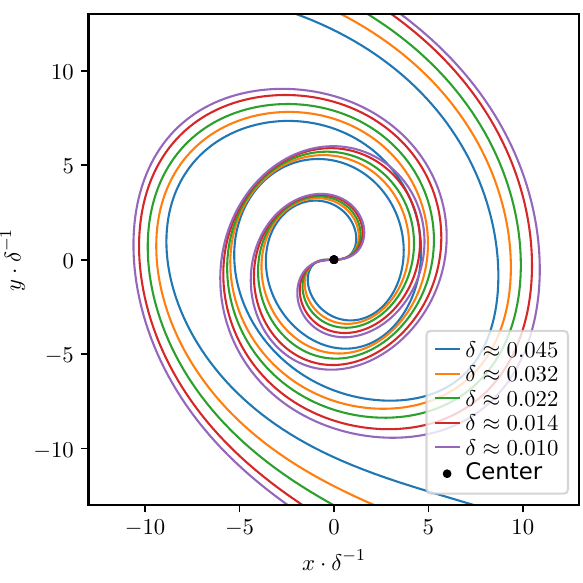}
        \caption{Re-scaled by $\delta^{-1}$, as done in \citep{Caflisch2022}}
    \end{subfigure}
    \caption{Spatial re-scaling with different factors 
    for the fourth time that $\partial\bm{x}(\theta, t)/\partial \theta\approx0$ in the $y$ direction for different $\delta$-values.}
    \label{fig:rescaling-spatial-comparison}
\end{figure}
This is also visible in figure~\ref{fig:vortex-turnover}. 
Here the occurrence of the vanishing gradient is plotted over re-scaled time, where each occurrence marks when the gradient of the material line at the spiral center becomes zero, i.e. when either the $x$- or $y$-component of the derivative of the material line position $\frac{\partial{\bm x}(\theta,t)}{\partial\theta}$ vanishes. As the vortex turnover over re-scaled time occurs in constant periods (linear growth in figure~\ref{fig:vortex-turnover}), the vortex rotates with constant speed. It is similar to a solid body rotation and portrays a stable vortex core. \\
Additionally, the palinstrophy growth was also found to be re-scalable in time using $t_s$ and $delta$, as shown in figure \ref{fig:rescaling-pal}. 
In all graphs of figures \ref{fig:rescaling-temporal-minmax} and \ref{fig:rescaling-temporal-2}, temporal scaling has been used. As for different $\delta$-values the results are similar, those four criteria were used to match them and determine an empirical relation. The ansatz was taken from \citet{Caflisch2022} using $(t - t_s)\delta^{-1}$, where $t_s$ is the singularity time and $\delta^{-1}$ as re-scaling factor. The flow scales linearly with decreasing $\delta$-values. The critical times are not equal to $t_s = 1.505$, the value of the Birkhoff--Rott equation but were observed to be slightly larger. They are closely constant for larger $\delta$-values but changed once it was decreased, possibly going towards the critical time of the BR-equation for vanishing vortex sheet thickness $\delta$. The linear scaling with $\delta^{-1}$ in comparison to the scaling of $\delta^{-2/3}$ for viscous flow reported by \citet{Caflisch2022} results from different flow dynamics. With no viscosity the whole dynamics of the condensed vortex core is an accumulation of vorticity present in the initial vortex sheets, which scales directly with the thickness $\delta$. \\
\begin{figure}
    \center
    \begin{subfigure}[t]{0.48\linewidth}
        \centering
        \includegraphics[width=\linewidth]{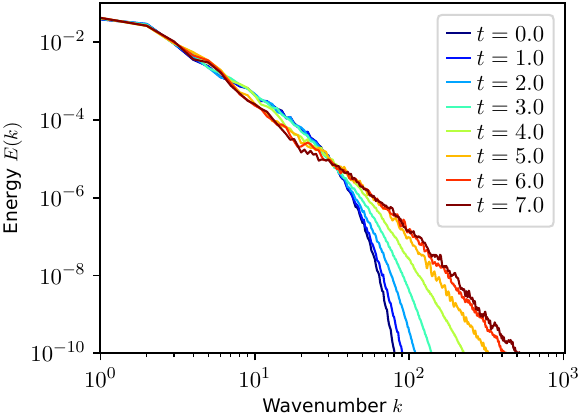}
        \caption{$\delta=0.045$}
        \label{fig:energy-spectra-d0.045}
    \end{subfigure}
    \hspace{0.2cm}
    \begin{subfigure}[t]{0.48\linewidth}
        \centering
        \includegraphics[width=\linewidth]{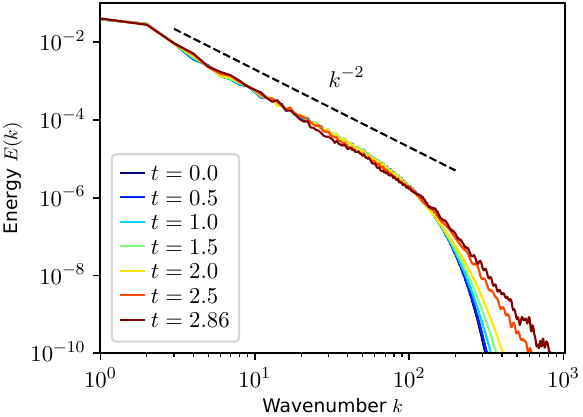}
        \caption{$\delta=0.01$}
        \label{fig:energy-spectra-d0.01}
    \end{subfigure}
    \caption{Energy spectra $E(k)$ defined in equation \ref{eqn:energy-spectra}
    for different $\delta$-values. Both start with an initial scaling of $k^{-2}$ with exponential decay for large $k$ as given in figure \ref{fig:IC-power-spectra}.}
    \label{fig:energy-spectra}
\end{figure}
The material curves for different $\delta$-values have strong spatial self-similarity. However, while \citet{Caflisch2022} reported a scaling of $\delta^{-1}$, this did not match our computed results. These show an empirical scaling of $\delta^{-0.9}$, which is slightly different, as shown in figure \ref{fig:rescaling-spatial-comparison}. The similarity scaling is valid for all reported simulations and all times around the vortex cores and only breaks down with the vortex merging process for later times (observed for $\delta\approx0.045$ for $t>7$). \\ %in advanced stage. \\
The characteristic of the initial energy spectra, defined in equation~\ref{eqn:energy-spectra}, have already been explained in section~\ref{sec:set-up}. For larger $\delta$-values the initial profile with exponential decay experiences a shift of energy to finer scales (figure \ref{fig:energy-spectra-d0.045}).
In comparison, the results with lower $\delta$-values with initially more pronounced $k^{-2}$ Dirac-scaling show only little shift in energy from larger to finer scales (figure \ref{fig:energy-spectra-d0.01}). However, for both simulations the two vortices have already emerged for the captured time and performed several windings, while the results for figure~\ref{fig:energy-spectra-d0.01} did not yet observe any developed distortion from the vortex merging process. Eventually, once this global process continues the energy  spectra are expected to behave similarly as to that for larger $\delta$-values. The formation of the local vortices exhibits therefore only little impact on the energy spectra. \\
The two vortices which form are quite unstable once perturbed to sufficient degree. With ongoing global merging process with increasing influence on the flow field, the shape of the vortex core becomes more and more elliptic (figure \ref{fig:prolonged-sim}). This brings an imbalance to the velocity field and secondary vortices form, breaking down the regularity. The individual parts of the spiral arm become compressed into filament structures and 
secondary vortices form. \\
\begin{figure}
    \center
    \begin{subfigure}[t]{0.28\linewidth}
        \centering
        \includegraphics[height=3.9cm]{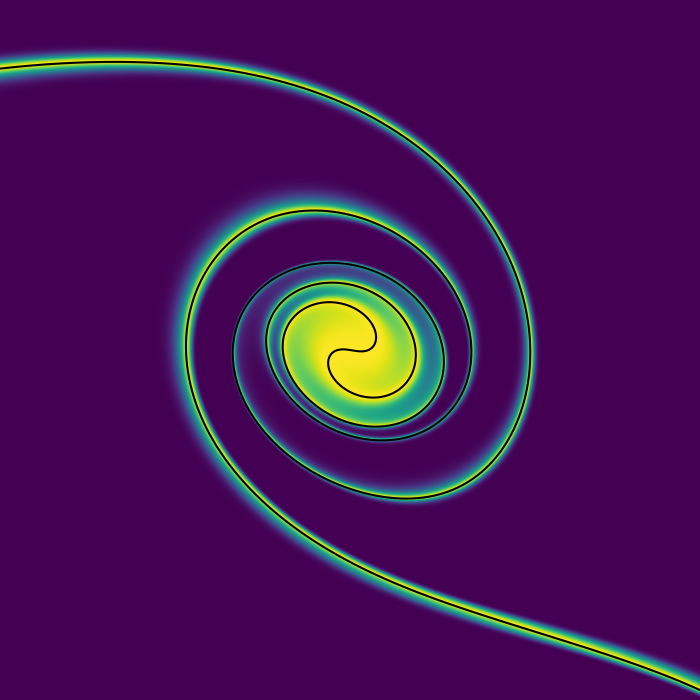}
        \caption{$t=5$}
    \end{subfigure}
    \hspace{0.2cm}
    \begin{subfigure}[t]{0.28\linewidth}
        \centering
        \includegraphics[height=3.9cm]{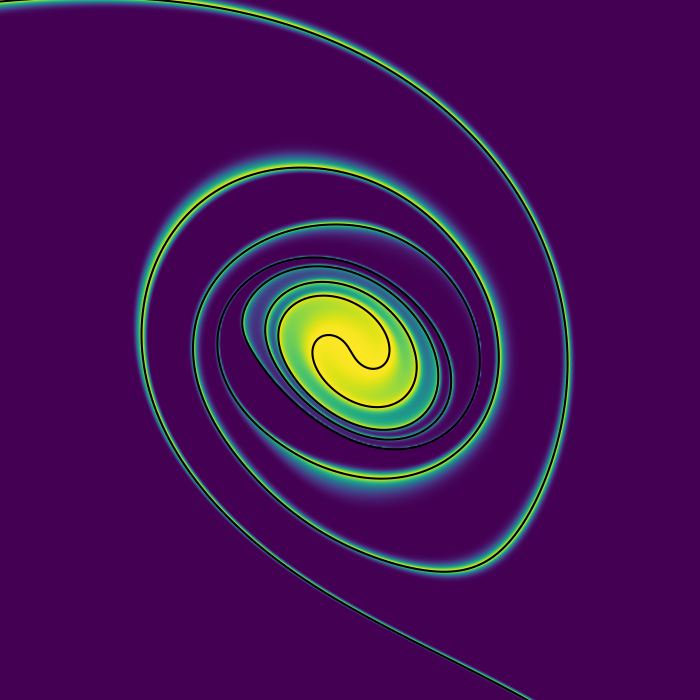}
        \caption{$t=6.2$}
    \end{subfigure}
    \hspace{0.2cm}
    \begin{subfigure}[t]{0.35\linewidth}
        \centering
        \includegraphics[height=3.9cm]{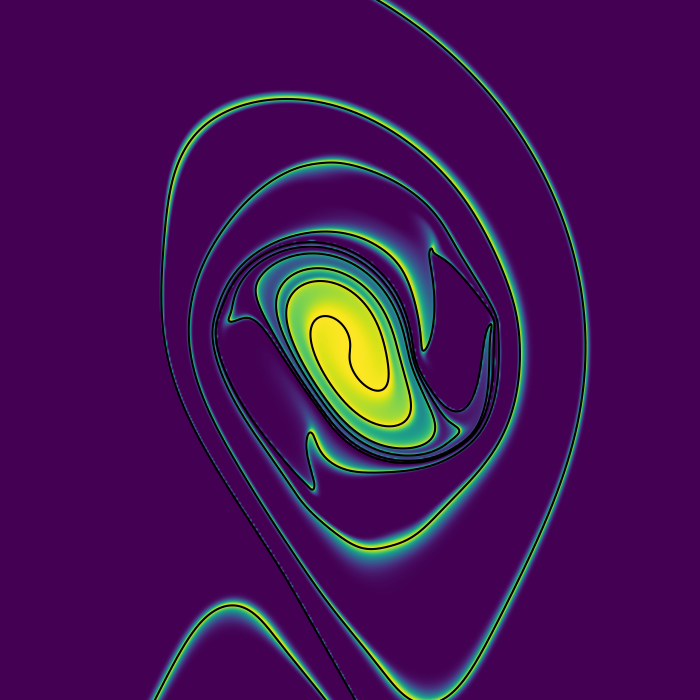}
        \includegraphics[height=3.9cm]{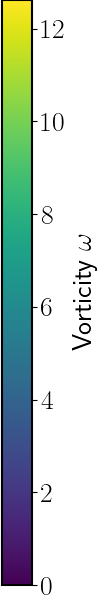}
        \caption{$t=7.4$}
    \end{subfigure}
    \caption{Material line (in black) and vorticity for $\delta=0.032$ at different times. The initially more stable spiral arms experience distortions and filament structures start to emerge with ongoing global vortex merger process.}
    \label{fig:prolonged-sim}
\end{figure}

%-----------------------------------------------------------------------------------
\section{Singularity analysis}
\label{sec:singularityanalysis}

The complex singularity of an analytic function can be analyzed using the Fourier transform. The width of the analyticity strip
can be obtained by considering the asymptotic behavior of the Fourier spectrum governed by Laplace's formula.
To obtain information about singularities outside the analyticity strip the Borel--Polya--van der Hoeven (BPH) method has been proposed \citep{PaulsFrisch2007}.

\subsection{Borel--Polya--van der Hoeven method}

The Borel--Polya--van der Hoeven (BPH) method \citep{PaulsFrisch2007} is a numerical tool for finding complex singularities of single variable functions by combining the information on singularities obtained from Borel transforms for Taylor series \citep{Polya1929} with numerical techniques for asymptotic interpolation \citep{Hoeven2006}. This method has been successfully applied to find singularities for the 1D Burgers equations \citep{PaulsFrisch2007} and has been used to analyze potential singularity formation in the 2D Euler vortex sheet problem \citep{Caflisch2022}. The following is a summary of the method along with some small modifications implemented to adapt to our case. 

Given a complex function $f(Z)$ with formal power series
\begin{equation} \label{eqn:def_f}
    f(Z) = \sum_{n=0}^\infty a_n Z^n,
\end{equation}
its Borel transform and Borel--Laplace transform are given respectively by
\begin{equation} \label{eqn:BL_transform}
    f^B(\xi) = \sum_{n=0}^\infty \frac{a_n}{n!} \xi^n, \quad \quad  f^{BL} (Z) = \frac{1}{Z} f \left( \frac{1}{Z} \right) =  \sum_{n=0}^\infty \frac{a_n}{Z^{n+1}},
\end{equation}
so named since $f^{BL}$ is formally the Laplace transform of $f^B$.

The BPH method is built on P\'{o}lya's theorem which is based on the observation that for $a_n = c^n$ for some complex number $c$, the Borel transform is the exponential
\begin{equation}
    f^B(\xi) = \sum_{n=0}^\infty \frac{(c\xi)^n}{n!} = e^{c \xi},
\end{equation}
while the Borel--Laplace transform is a simple pole at $c$:
\begin{equation}
    f^{BL}(Z) = \sum_{n=0}^\infty \frac{c^n}{Z^{n+1}} = \frac{1}{Z-c}.
\end{equation}

Writing $f^B$ in polar coordinates using $\xi = r e^{-i \theta}$ (reversed phase parametrization for convenience), and $c = |c| e^{i \phi}$, we have that 
\begin{equation}
    \ln (f^B(r, \theta) ) = |c| r (\cos(\phi - \theta) +i \sin(\phi-\theta) ),
\end{equation}
and therefore $\partial_r \ln ( |f^B(r, \theta)| )$ is maximized at $\theta = \phi$ to value $|c|$ thereby revealing the position of the pole. 

For $f^{BL}$ given by a linear combination of multiple isolated poles $f^{BL}(Z) = \frac{C_1}{Z-c_1} + \frac{C_2}{Z-c_2} + \dots + \frac{C_m}{Z-c_m}$, the supporting function $\sigma (\theta)$ is given by
\begin{equation}
    \sigma (\theta) := \lim_{r \to \infty} \partial_r \ln ( |f^B(r, \theta)| ) = \max_{j = 1, 2, \ldots, m} |c_j| \cos(\phi_j - \theta) .
\end{equation}
The curve $\sigma(\theta) e^{i \theta}$ then describes the convex hull of the set of singularities. This also means that only the  singularities of $f^{BL}$ furthest from the origin are identified from the vertices of the convex hull. Since the Borel--Laplace transform $f \mapsto f^{BL}$ involves the change of variable $Z \mapsto 1/Z$, we have that the complex reciprocals of these singularities are the singularities of $f$ closest to the origin.

This method is then applied to estimate the location of complex plane singularities for a real valued periodic function $u(x)$ on $S^1 \sim [-\pi, \pi)$. We write $u$ in terms of Fourier series and make the analytic extension in some strip around the real axis
\begin{equation}
    u(z) = \sum_{k \in \mathbb{Z}} \hat{u}_k e^{ikz}
\end{equation}
which is decomposed as the sum of two functions
\begin{subequations}
    \begin{align}
        & u^+ (z) = \sum_{k > 0} \hat{u}_k e^{ikz} , \\
        & u^- (z) = \sum_{k > 0} \hat{u}_{-k} e^{-ikz} .
    \end{align}
\end{subequations}
We note that $u^+$ is analytic in the upper half plane, i.e. poles are contained in the lower half plane. Similarly, all poles of $u^-$ are in the upper half plane. The change of variables $Z = e^{-z}$ and $Z = e^{-iz}$ are made to $u^+$ and $u^-$ respectively to yield
\begin{subequations}
    \begin{align}
        & f_1 (Z) = u^+ ( -i \ln Z ) , \\
        & f_2 (Z) = u^- ( i \ln Z ) ,
    \end{align}
\end{subequations}
where we use the $[-\pi, \pi)$ branch cut. We note now that both $f_1$ and $f_2$ are analytic inside the unit disk corresponding to the image of the upper and respectively lower half planes under the change of coordinate. Applying the BPH method on $f_1$ and $f_2$ then reveals the poles closest to the unit circle and hence closest to the real axis after the logarithmic change of coordinate.

The algorithm of the BPH method is summarized as follows. For a function $f$ given as a truncated power series in \eqref{eqn:def_f} with coefficients $a_n$, we compute its Borel transform $f^B(\xi)$ on a grid in polar coordinates given by $\xi_{j, k} = r_j e^{-i \theta_k}$. The grid points are given by
\begin{equation}
    r_j = r_0 + \frac{j R}{M}, \quad \quad \theta_k = \frac{2 \pi k}{K}
\end{equation}
for $j = 0, 1, 2, \ldots, M-1$ and $k = 0, 1, 2, \ldots, K-1$ with an appropriate choice of $r_0$ and $R$ to be discussed later.

At each grid point, the Borel transform is given by a $N$-truncated sum of \eqref{eqn:BL_transform}. If we select the number of rays $K$ to be equal to the number of power series terms $N$, we get that
\begin{equation}
    f^B(\xi_{j, k}) = \sum_{n=0}^{N-1} a_n  \frac{r_j^n }{n!} e^{-i\frac{2\pi k n}{N}} = \sum_{n=0}^{N-1} a_n  \frac{r_j^n }{n!} e^{-i \theta_k n} ,
\end{equation}
which, for fixed $j$, is the expression for the discrete Fourier transforms of the sequence $\{b_n\}_{n=0}^{N-1}$ with
\begin{equation} \label{eqn:BFFT_terms}
    b_n = a_n \frac{r_j^n }{n!} = a_n \left( r_j \exp \left(\frac{- \ln \Gamma (n+1) }{n} \right) \right)^n = \exp \left( n \ln r_j + \ln a_n - \ln \Gamma (n+1)  \right).
\end{equation}
FFT algorithms are used to compute the discrete Fourier transform in order to reduce round-off errors and improve speed and the log-Gamma function is used to avoid numerical overflow in the computation of the larger $n$ terms.

Then for each fixed $\theta_k$, $\sigma(\theta_k)$ can be estimated from finite difference on $\ln ( |f^B(r, \theta)| )$. Higher order asymptotic interpolation could be used but would require high precision arithmetic. For double precision computations, we opted for simple finite difference. From the supporting function $\sigma(\theta)$, we can identify $S^{BL} = \{ y^*_1, y^*_2, \ldots, y^*_s \}$, the vertices of the convex hull of the singularities of $f^{BL}$ which we recall lie within the unit complex disk. The choice of the values of $r_0$ and $R$ needs to be adjusted carefully. Since along the ray of each $y_i^*$, the Borel transform is expected to grow as $f^{B} (\xi) \sim \exp(y_i^* \xi )$, in order to capture the exponential growth, one should pick $r_0$ larger than $1/|y_i^*|$. The maximum radius $R$ should be picked as large as possible until the effects of finite precision arithmetic start dominating.

Applying the inverse Borel--Laplace transform gives us the set of singularities $S = \{ Z^*_1, Z^*_2, \ldots, Z^*_s \}$ for $f$, where $Z^*_j = 1/y_j^*$, which are now the singularities of $f$ closest to the unit circle. The locations of the singularities $z^*_i$ of $u^\pm$ are then obtained by performing the inverse transform $z^*_i = \mp i \ln Z $. This means that if one has an \emph{a priori} estimate for the radius of analyticity $\delta$ of $u$, the value $r_0$ should be chosen larger than $e^\delta$.

\begin{figure}
    \centering
    \begin{subfigure}[t]{0.47\linewidth}
        \centering
        \includegraphics[width=1\linewidth]{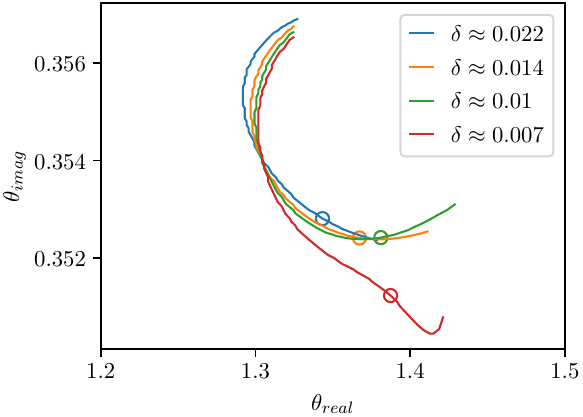}
        \caption{}
    \end{subfigure}
    \hspace{0.2cm}
    \begin{subfigure}[t]{0.47\linewidth}
        \centering
        \includegraphics[width=1\linewidth]{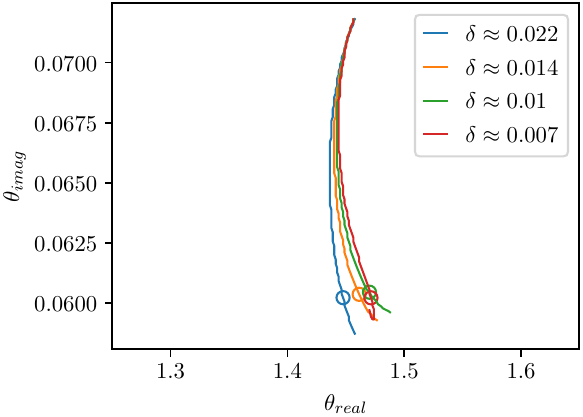}
        \caption{}
    \end{subfigure}
    \caption{Evolution of the complex plane singularities for the true vortex strength (a) and curvature (b) for various values of $\delta$ from $t=0.8$ to $1.6$. The position of the singularity at time $t=1.5$ is marked by the circles. The singularity positions are computed using the BPH method using radii in $[1, 2.4]$. The use of larger radii was not possible due to the presence of noise in the data.}
    \label{fig:singularities}
\end{figure}

We applied the BPH method to analyze the singularity formation in the vortex core curve $\gamma$. As in \citep{Caflisch2022}, we compute the closest complex singularities for the vortex strength function $\sigma$ and the curvature function $\kappa$ given by
\begin{align}
    & \sigma = | \dot{\gamma}  |^{-1} ,\\
    & \kappa  = \frac{ | \dot{\gamma} \times \ddot{\gamma}  | } { | \dot{\gamma}  |^{3} } .
\end{align}
Since the CMM relies on Hermite cubic interpolation for the stream function which yields a $C^0$ velocity field with small discontinuities in the derivative (order of $h^3$ where here $h \sim 1e-9$), the evolved vortex curve $\gamma$ is also only $C^0$. The lack of regularity shows up as noise in the large wave-numbers of $\hat{\gamma}$. To perform the BPH analysis on this data, a smoothing kernel $\exp ( -1\times 10^{-7} k^4 )$ in Fourier space and a truncation to the 10000 first wave-numbers is applied to the curve as pre-processing. The results are shown in figure \ref{fig:singularities}, where the position of the complex singularities are drawn for times between $0.8$ and $1.6$. The general trajectories of the complex plane singularities before time $t=1.5$ are in agreement with the results found for the viscous case in \citep{Caflisch2022}. We observe that all singularities move towards the real axis as time advances until $t=1.5$ (marked with circles) after which they seem to move away in some cases. The meaning of this is not clear from these tests as the accuracy of the singularity analysis is very limited due to the limited available arithmetic precision. We cannot exclude the possibility that the BPH method is detecting another complex singularity which overtakes the one tracked in the above figures resulting in a transition to a new singularity which is closest to the real axis.

%-----------------------------------------------------------------------------------
\section{Conclusions}
\label{sec:concl}

The flow of vortex layers governed by the incompressible 2D Euler equations has been computed for successively decreasing layer thickness using the characteristic mapping method. This semi-Lagrangian method features exponential resolution in linear time and thus allows to capture the exponential growth of the vorticity gradients. Our results agree with pseudo-spectral computations of \citet{Caflisch2022} and go even beyond their reported results.
In particular, the range of results for vanishing vortex sheets thickness was extended down to $\delta\approx0.007$ and longer captured simulations enabled further analysis of the emerging vortex structures.
Energy and enstrophy are conserved to a high degree thanks to the non-dissipative feature of the CM method. The palinstrophy shows at the beginning a super-exponential growth, which is intercepted and dominated by a stronger exponential growth once the two vortices form.
The dynamics of the center-line of the vortex sheet can be described by measures of the arc length, curvature and vortex strength.
The arc length and vortex strength showcase, that the emerging vortices have a region of very strong compression around the vortex centers and of elongation in the spiral arms that form. These form a peak in compression directly at the center and scale with decreasing vortex sheet thickness $\delta$.
For the limit of $\delta \rightarrow 0$ the vortex strength forms a singularity at the center of rotation where it is compressed into it with vortex strength tending to infinity.
The curvature displays similar singular behavior. Instead of forming maximum values directly at the vortex center, it forms two at the edge of the condensed vortex blobs. These are of opposite sign and describe the transition from the vortex blob formation to the spiral arms and additionally scale with vortex sheet thickness $\delta$, again going to infinity for the limit of vanishing $\delta$-values, further supporting the formation of singularity for non-smooth initial data suggested in \citet{Caflisch2022}.
These quantities also show strong self-similarity in time and space. With the help of the curvature, vortex strength and turnover time, the temporal rescaling of the vortex dynamics was unveiled. In fact, after initial build up these vortices rotate with constant speed, similar to a solid body rotation. The maximum values for the curvature and vortex strength also do not observe monotonous growth over time, but after a steep increase they oscillate around a common value. The temporal rescaling was not found to be consistent with that of the BR-equation and similarities are clearly present.
In space, the formed vortices show strong self-similarity even for many turnover events and over a large range of $\delta$-values. The observed scaling slightly differs from that found in \citet{Caflisch2022} in the viscous case, i.e. for Navier--Stokes.
The energy spectra show for small $\delta$ values a power law scaling with slope close to $-2$. 
Simulations for longer times show, that at some point the vortex merger starts to distort the round-shaped vortex blobs, this leads to instability in the roll-up process and secondary vortices start to form, eventually breaking down the spiral structure by continuous filamentation.

\medskip
In future work we will apply CMM to compute fine scale structures of vorticity gradients in 2D Euler. The transport equation of the vorticty gradients contains a source term for vorticity gradient stretching, similar to the vortex stretching source term in 3D Euler. The latter can be solved likewise by CMM, similar to what has proposed in \citet{CMM3D}.

Another challenging perspective is applying CMM to study Euler flows in 3D and to investigate numerically possible singularities. First results presenting low resolution computations for the \citet{Kerr1993, HoLi2007} initial condition can be found in \citet{CMM3D}. High resolution 3D  computations considering different flow configurations, like \citet{Kerr1993, HoLi2007} and more recently the one by \citet{Moffatt2019, Moffatt2020}, will be published in forthcoming work.

\bigskip

\section*{Acknowledgements}
The authors acknowledge partial funding from the Agence Nationale de la Recherche (ANR), grant ANR-20-CE46-0010-01. This work was granted access to the HPC resources of IDRIS under the
allocation 2022-91664 made by GENCI.
Center de Calcul Intensif d’Aix-Marseille is acknowledged for granting access to its high performance computing resources.
Part of this research was finalized while KS was visiting the Institute for Mathematical and Statistical Innovation (IMSI) at University of Chicago, which is supported by the National Science Foundation (Grant No. DMS-1929348). JCN Acknowledges partial support from the NSERC Discovery Program.  

\bigskip

\hrule

\bigskip

\bigskip

%------------------- appendix ----------------
\appendix{\bf \Large Appendix}

\medskip

In the following we present numerical validation of the open access CMM cuda code~\citep{url_cmm} similar to what has been done in \citet{CMM2D} for the Matlab implementation.

\subsection*{Validation of convergence order in space and time for Cuda-code \citep{url_cmm}} \label{sec:conv-order}

\bigskip

All convergence tests presented use the same parameters as in
%here have been done similar to 
\citet{CMM2D} in order to make the results comparable. 
The 4-mode-flow was used as initial condition with,
\begin{align}
    \omega_0(x,y) = \cos(x) + \cos(y) + 0.6 \cos(2x) + 0.2 \cos(3x) \; .
\end{align}
In table \ref{tab:ref-settings}, all numerical parameters for the reference simulations can be found for which the error analyses have been carried out. The grid for the flow map $\chi$ ($N_{coarse}$), initial vorticity $\omega_0$ ($N_{fine}$) and sampling of the vorticity for the FFT ($N_\omega)$ were chosen of same size and the velocity field uses a grid with increased values to reduce the influence of the non-smoothness of the velocity-gradients. The time-step $\Delta t$ is set to a Courant--Friedrich--Lewis (CFL) number of $2$. The size $\epsilon_m$ defines the stencil size for the GALS method. In the presented computational study no low-pass filtering and no remapping was used in order to capture the sub-map error correctly. All simulations were run until a final time of $t=1$. For further understanding of the impact of parameters, the readers are directed to \citep{CMM2D, Bergmann2022}.

\begin{table}[H]
    \centering
    \begin{tabular}{ c | c || c | c}
        Name & Value & Name & Value \\ \hline
        $N_{\text{coarse}}$ & $1024$ & $N_{\text{fine}}$ & $1024$ \\
        $N_{\psi}$ & $2048$ & $N_{\omega}$ & $1024$ \\
        $h_{\text{fluid}}$ & $1/512$ & & \\
        $\epsilon_m$ & $10^{-3}$ & Initial condition & 4-mode-flow \\
        Fluid time scheme & RK3 & Map update stencil & 4th order
    \end{tabular}
    \caption{Settings of the reference simulation~\citep{CMM2D}. \label{tab:ref-settings}}
\end{table}

The errors of four quantities were computed to examine the convergence order. Those are the flow map and vorticity error in $L_\infty$-norm and the energy and enstrophy conservation error in $L_2$-norm. All the errors were evaluated by sampling the map on a uniform $2048^2$-grid and computing the vorticity and velocity respectively on this map.
\begin{align}
    \text{Map error} &= || \chi_{ref}(\cdot, t_n) - \chi(\cdot, t_n) ||_\infty \label{eqn:err_map} \\
    \text{Vorticity error} &= || \omega_{ref}(\cdot, t_n) - \omega(\cdot, t_n) ||_\infty \label{eqn:err_vort} \\
    \text{Energy error} &= || \omega(\cdot, t_n)^2||_2 - ||\omega(\cdot, t_0)^2 ||_2 \label{eqn:err_en} \\
    \text{Enstrophy error} &= || \bm{u}(\cdot, t_n)^2||_2 - ||\bm{u}(\cdot, t_0)^2 ||_2 \label{eqn:err_enst}
\end{align}
According to \citet{CMM2D}, an error bound for the characteristic map is given by
\begin{align}
    \tilde{\mathcal{E}}^n = \mathcal{O} \left( \Delta x ^2 \min(\Delta t, \Delta x^2  \Delta t^{-1}) + \Delta t ^s + \Delta t ^p \right)
\end{align}

\begin{figure}
	\centering
    \begin{subfigure}[t]{0.3\linewidth}
        \centering
        \includegraphics[width=\linewidth]{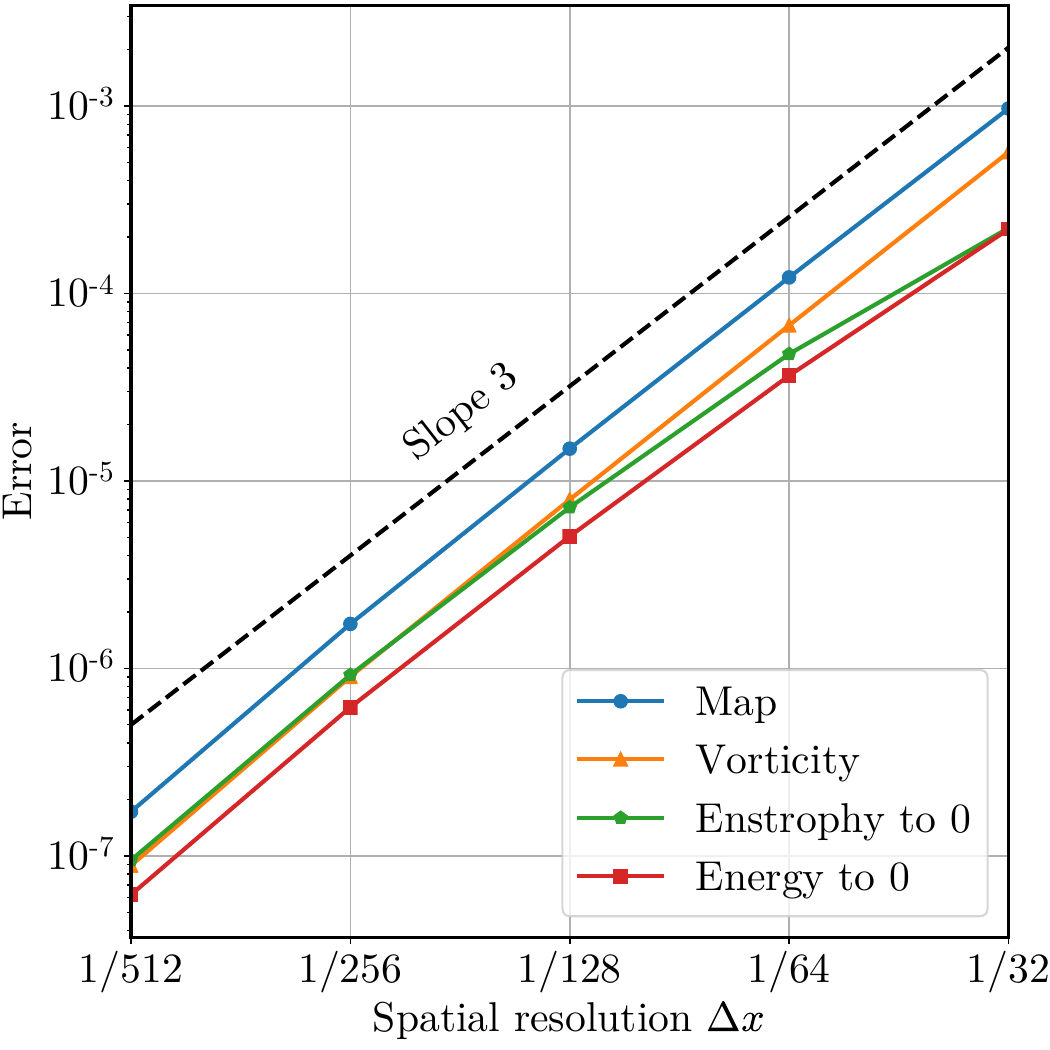}
        \caption{\centering Convergence errors in space for fluid flow}
    \end{subfigure}
    \hspace{0.2cm}
    \begin{subfigure}[t]{0.3\linewidth}
        \centering
        \includegraphics[width=\linewidth]{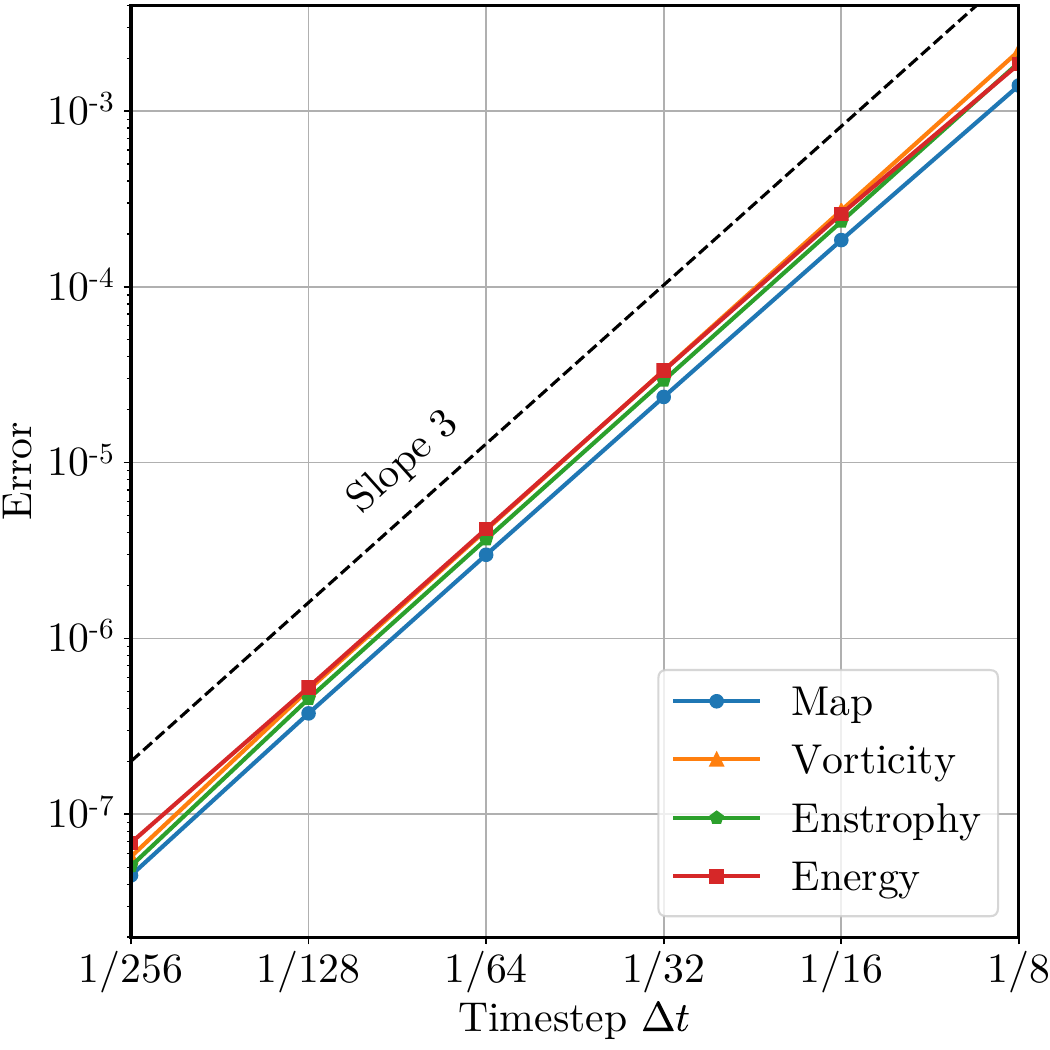}
        \caption{\centering Convergence errors in time for fluid flow}
    \end{subfigure}
    \hspace{0.2cm}
    \begin{subfigure}[t]{0.3\linewidth}
        \centering
        \includegraphics[width=\linewidth]{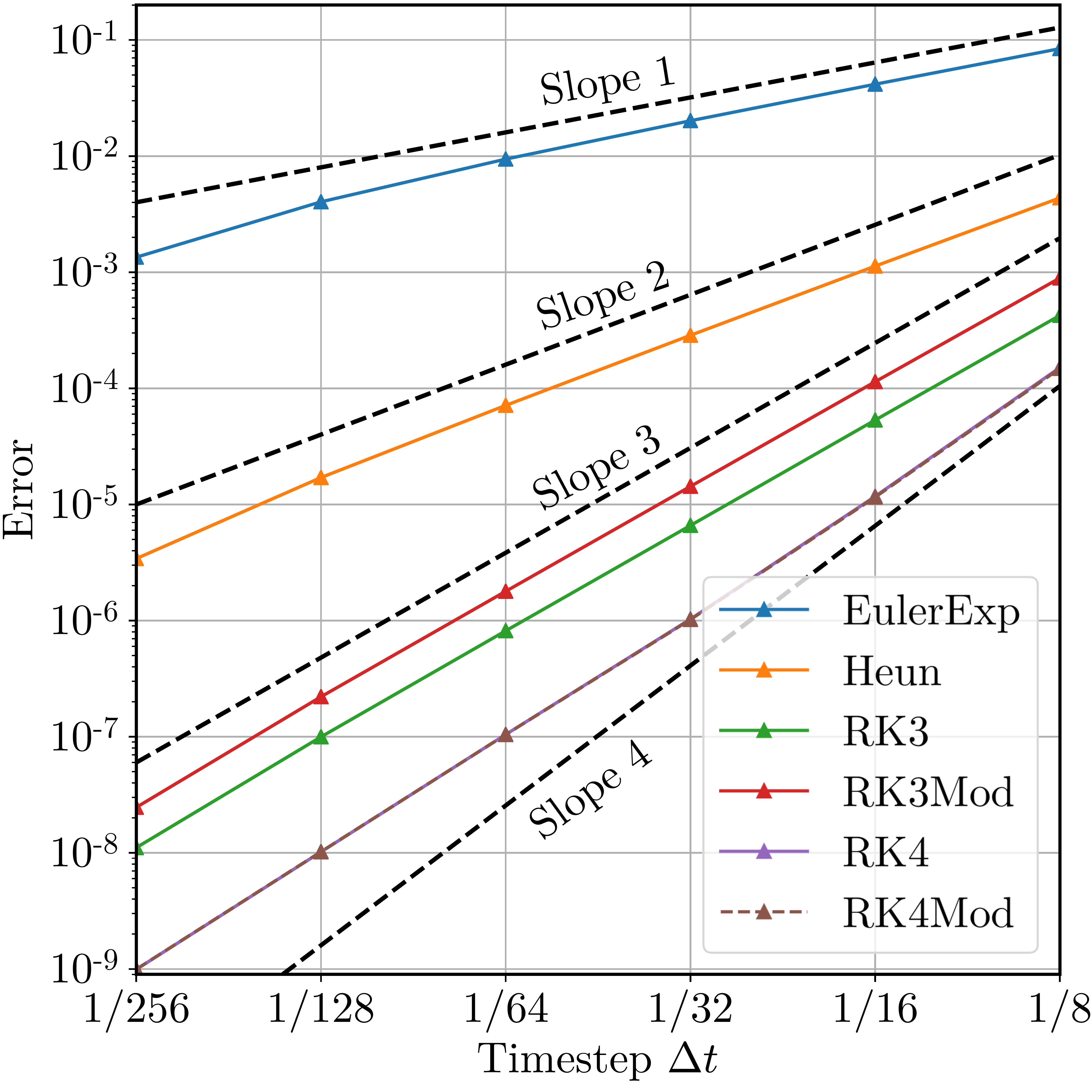}
        \caption{\centering Convergence errors in time for fluid particles}
        \label{fig:inv-conv-part}
    \end{subfigure}
	\caption{Convergence errors from equation \ref{eqn:err_map}-\ref{eqn:err_enst} in space and time for the flow variables and equation \ref{eqn:err_part} for the fluid particles. Expected convergence rates are plotted with dotted lines.}
	\label{fig:inv-conv}
\end{figure}

Here $s$ and $p$ are the orders  of the s-stage Runge--Kutta scheme and the used order of Lagrange time interpolation for the velocity, respectively. Values were set as $p=s$ to balance computation requirements with achieved numerical accuracy.
Due to the bi-cubic spatial Hermite interpolation used, errors of order $\mathcal{O}(\Delta x^3)$ for the convergence order in space are expected (figure \ref{fig:inv-conv}). One order is reduced to the theoretical 4th order as the velocity to advect the map using the GALS framework is sampled as a first order derivative from the stream function. The Cuda code is capable of including first to fourth order Runge--Kutta scheme. An example for convergence in time with third order scheme paired with third order Lagrange-interpolation of the velocity is depicted in figure \ref{fig:inv-conv}. Again, all quantities converge with the expected order.

The evolution of fluid particles used in the manuscript to track the material line of the vortex sheet, is computed using Lagrangian point particles, where the velocity is set to that of the fluid flow, %$\bm{u}_f$:
\begin{align}
    \frac{d \bm{x}_{p}}{dt} &= \bm{u} \label{eqn:part-x}
\end{align}
Here $\bm{x}_{p}$ is the position of the point particle and $\bm{u}$ the fluid velocity. For time integration we tested different numerical schemes (cf. figure~\ref{fig:inv-conv-part}), for spatial interpolation of the velocity at the particle positions bi-cubic Hermite interpolation is applied.
Similarly to the map error quantities, the particle position error can be defined as
\begin{align}
    \text{Particle error} = || \bm{x}_{p,ref}(\cdot, t_n) - \bm{x}_p(\cdot, t_n) ||_\infty \label{eqn:err_part}
\end{align}
The particles where initially scattered with uniform random distribution over the computational domain. Due to the volume preservation property of the incompressible flow, their distribution will remain uniform as well. All fluid parameters were kept similarly to the reference computation (table \ref{tab:ref-settings}) and $10^6$ particles were deployed. The tested time-stepping methods show the expected convergence orders. Merely fourth order deployed time-stepping schemes are reduced to third order, which is due to the third order convergence of the fluid velocity (figure \ref{fig:inv-conv-part}).

\bibliographystyle{jfm}
%\bibliography{jfm}

\end{document}